\documentclass[acmsmall,screen]{acmart}

\startPage{1}
\setcopyright{none}
\renewcommand\footnotetextcopyrightpermission[1]{}
\usepackage{booktabs}   
\usepackage{subcaption} 
\usepackage{makecell}
\usepackage{multirow}
\usepackage{xcolor}
\usepackage{soul}
\usepackage{colortbl}
\usepackage{hyperref}
\usepackage{enumitem}
\usepackage{algorithm}
\usepackage{algorithmicx}
\usepackage[noend]{algpseudocode}
\usepackage{graphicx}
\usepackage{xcolor}
\usepackage{booktabs}
\usepackage{subcaption}
\usepackage{listings}
\usepackage{mathpartir}
\usepackage{adjustbox}
\usepackage{xspace}
\usepackage{cleveref}

\makeatletter 
\newcount\SOUL@minus
\makeatother  

\newcommand{\tuple}[1]{\ensuremath{\langle #1 \rangle }}
\newcommand{\mtext}[1]{\ensuremath{\text{\it #1}}}
\newcommand{\lamirage}{\textsc{LaMirage}\xspace}
\newcommand{\bifi}{\textsc{BiFi}\xspace}
\newcommand{\codex}{\textsc{Codex}\xspace}
\newcommand{\codexedit}{\textsc{Codex-Edit}\xspace}
\newcommand{\codebert}{\textsc{CodeBERT}\xspace}
\newcommand{\ignore}[1]{{}}
\newcommand{\grmtools}{\textsc{grmtools}\xspace}
\newcommand{\exceldesktop}{\textsc{Excel-Desktop}\xspace}

\setlist{nolistsep,leftmargin=*}

\DeclareMathOperator{\argmax}{argmax}

\DeclareMathOperator{\dist}{dist}

\DeclareRobustCommand{\hlcyan}[1]{{\sethlcolor{lime}\hl{#1}}}
\DeclareRobustCommand{\hlpink}[1]{{\sethlcolor{pink}\hl{#1}}}
\definecolor{ForestGreen}{RGB}{34,139,34}

\newcommand{\SymAssign}[0]{\ensuremath{\leftarrow}}
\algnewcommand\algorithmicforeach{\textbf{for each}}
\algdef{S}[FOR]{ForEach}[1]{\algorithmicforeach\ #1\ \algorithmicdo}

\algtext*{EndWhile}
\algtext*{EndIf}
\algtext*{EndFor}

\newcommand\cldots{\makebox[1em][c]{.\hfil.\hfil.}\thinspace}
\def\sfixed{{\hat{s}}}

\newcommand{\cheapsubsection}[1]{\vspace{0.17cm}\noindent\textbf{#1}}
\newcommand{\cheapsubsectionit}[1]{\vspace{0.17cm}\noindent\textit{#1}}

\begin{document}

\title{Neurosymbolic Repair for Low-Code Formula Languages}


\author[R. Bavishi]{Rohan Bavishi}
\authornote{Work done during an internship with the PROSE team at Microsoft} \authornote{Equal contribution}
\affiliation{
  \institution{UC Berkeley}
  \country{USA}
}
\email{rbavishi@cs.berkeley.edu} 

\author[H. Joshi]{Harshit Joshi$^\dagger$}
\affiliation{
  \institution{Microsoft}
  \country{India}
}
\email{t-hjoshi@microsoft.com} 

\author[J. Cambronero]{Jos{\'e} Cambronero}
\authornote{Authors in alphabetic order} 
\affiliation{
  \institution{Microsoft}
  \country{USA}
}
\email{jcambronero@microsoft.com} 

\author[A. Fariha]{Anna Fariha$^\ddagger$}
\affiliation{
  \institution{Microsoft}
    \country{USA}
}
\email{annafariha@microsoft.com} 

\author[S. Gulwani]{Sumit Gulwani$^\ddagger$}
\affiliation{
  \institution{Microsoft}
    \country{USA}
}
\email{sumitg@microsoft.com} 

\author[V. Le]{Vu Le$^\ddagger$}
\affiliation{
  \institution{Microsoft}   
    \country{USA}
}
\email{levu@microsoft.com} 

\author[I. Radicek]{Ivan Radicek$^\ddagger$}
\affiliation{
  \institution{Microsoft}   
    \country{Croatia}
}
\email{ivradice@microsoft.com} 

\author[A. Tiwari]{Ashish Tiwari$^\ddagger$}
\affiliation{
  \institution{Microsoft}
    \country{USA}
}
\email{astiwar@microsoft.com} 

\begin{abstract}

Most users of low-code platforms, such as Excel and PowerApps,
write programs in domain-specific formula languages to carry out
nontrivial tasks. Often users can write most of the program they want,
but introduce small mistakes that yield broken formulas. These mistakes,
which can be both syntactic and semantic, are hard for low-code
users to identify and fix, even though they can be resolved with just a
few edits. We formalize the problem of producing such
edits as the \emph{last-mile repair} problem. To address this problem,
we developed \lamirage,  a LAst-MIle RepAir-engine GEnerator that combines
symbolic and neural techniques to perform last-mile repair in low-code formula languages.
\lamirage takes a grammar and a set of domain-specific constraints/rules, which jointly
approximate the target language,
and uses these to generate a repair engine that can fix formulas in
that language. To tackle the challenges of localizing the errors and ranking the candidate repairs, \lamirage
leverages neural techniques, whereas it relies on symbolic methods to generate candidate repairs. This combination allows \lamirage to find repairs that satisfy the provided grammar and constraints, and then pick the most natural repair. We compare \lamirage to state-of-the-art
neural and symbolic approaches on 400 real Excel
and PowerFx formulas, where \lamirage outperforms all baselines.
We release these benchmarks to encourage subsequent work in low-code
domains.
\end{abstract}




\maketitle

\section{Introduction}

Low-code (LC) platforms allow end users to build applications or carry out
complex calculations with little-to-no programming experience. These platforms
promise to democratize the access to computational tools and skills for
end-users across a wide range of domains. LC domains include traditional
end-user applications, such as spreadsheets (Excel~\cite{excel} and Google
Sheets~\cite{sheets}), but are increasingly expanding to more diverse areas,
such as robotic process automation frameworks (Power
Automate~\cite{powerautomate}, UIPath~\cite{uipath}), and enterprise apps (Power
Apps~\cite{powerapps}, Appian~\cite{appian}). While these tools usually offer a
graphical user interface with basic functionality, most nontrivial applications
built in this domain require the end user to write ``small programs'' or
\textit{formulas}. These formula-like languages are specifically designed for LC
users~\cite{powerfx-overview} and are tailored to mirror spreadsheet formula
languages, such as that in Excel, with which many users are already familiar.
However, low-code languages can support functionalities that go beyond
traditional spreadsheet-like languages. For example, formulas written in
PowerApp's PowerFx language can interact with the user interface.

Many of these platforms have significant user bases, resulting in a huge amount of code written as formulas. For example, Excel has 1 billion users~\cite{morgan-stanley-call}. Power Apps, an LC platform for building enterprise apps, is one of Microsoft's fastest-growing product offerings (based on a recent earnings call). Economically, the LC sector is expected to continue to grow substantially. For example, Gartner (a market consulting firm) predicts that up to 65\% of application development will be done on such platforms by 2024~\cite{venturebeat}.

Traditional software engineers have benefited from decades of academic research at the
intersection of programming languages, software engineering, and artificial intelligence. For example, program synthesis research has enabled 
engineers to quickly extract information
from complex logs~\cite{raza2017automated}, fix network policies~\cite{hallahan2017automated}, and wrangle
dataframes~\cite{bavishi2019autopandas}. Code search techniques, often integrated into
version control platforms such as GitHub, allow developers to quickly search for related
code snippets, often overcoming syntax-level differences~\cite{yogo}. Automated refactoring tools~\cite{bluepencil} facilitate tasks such as updating APIs~\cite{apifix}.
Techniques can automatically produce patches
by leveraging static analyzers, test suites, or 
code examples~\cite{goues2019automated}.

\begin{figure}[t]
    \includegraphics[width=0.9\columnwidth]{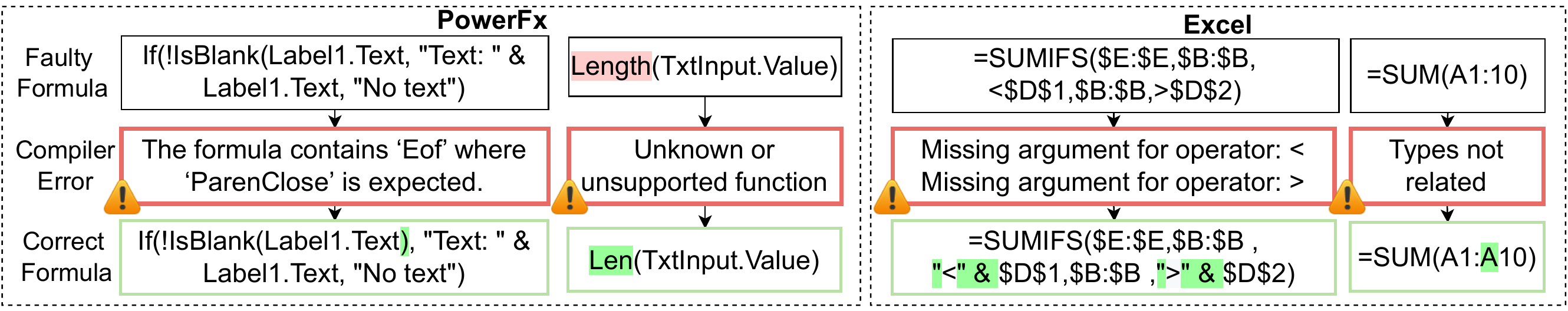}
    \vspace{-2.5mm}
    \caption{Real faulty formulas taken from help forums, their compile-error messages, and the correct formulas}
    \vspace{-2.5mm}
    \label{fig:examples}
\end{figure}

However, many of these advances have been focused on general purpose programming languages used by traditional programmers, such as Java, C\#, C++, or C, but not on LC platforms. This lack of LC-developer assistance can limit the accessibility of LC platforms, despite their intended goal of democratizing computational power. 
Our goal is to provide this new class of programmers with first-class feedback and tooling
comparable to that available for traditional programmers. To lead this effort, we 
first identify: \emph{where is the first place LC users tend to get stuck?} Based on 
user forums and discussions with both the Excel and PowerApps teams at Microsoft, we identified that small mistakes (first focused
on syntax, and then on semantics) paired with lack of error assistance are often the first stumbling block for LC programmers.
Figure~\ref{fig:examples} shows some user errors taken from help
forums~\cite{paforum, mrexcel} and the corresponding unhelpful compile-error
messages. This problem is compounded in the end-user domain as the formula authors often lack the experience to identify the correct repair even if the application were to provide a more detailed error message, which can lead to substantial frustration. Prior work studying programming novices has similarly found that syntax errors can contribute substantially to user frustration~\cite{drosos2017happyface}.

\noindent \paragraph{\textbf{Last-mile repair problem}} We studied faulty formulas
reported in LC help forums~\cite{paforum, mrexcel} and shared by the Excel and PowerApps teams at Microsoft.
We observed that: (1)~the formulas are often almost correct, (2)~most of the essential
components of the formulas are present in the correct order, (3)~the errors are
usually in terminal symbols, typically \emph{punctuation}, such as unbalanced
parentheses, missing commas, missing quotes, etc., and (4)~the repaired formulas are
usually within a small \textit{token edit distance}\footnote{Similar to string edit
distance, token edit distance involves adding, removing, or replacing characters, over tokens.}
of the intended correct formulas. Repair of such faulty formulas, being close to the correct formula, involve edits that an experienced programmer can identify without additional information, albeit with additional time and effort.
We focus on the problem of repairing such almost-correct formulas and call it the
\emph{last-mile repair problem} (\Cref{sec:problem}).

\paragraph{\textbf{Neural techniques are not enough:}}
Recently, the machine-learning (ML) community has taken up the
problem of repairing program errors in general-purpose languages 
by using deep neural networks~\cite{DeepFix, DrRepair, BIFI}.
Neural techniques, however, suffer from multiple shortcomings in our setting. 
First, these techniques are data-hungry and
require availability of huge program corpora to learn from. However, public corpora are not readily available in LC domains. 
Second, even large pre-trained models often struggle
to guarantee \emph{correctness} of the generated code and end up generating code that contain mistakes (including syntax errors), as shown by a recent work~\cite{synchromesh, austin2021program}. In the context of repair, it is critical that we do not introduce additional mistakes or ignore
existing ones that we can identify.

\paragraph{\textbf{Symbolic techniques are not enough:}}
Symbolic techniques, such as error recovery in parsers~\cite{Aho72,
FischerRecovery, NicolaeRajasekara14},
can provide some guarantees (e.g., syntactic validity), but their design is typically constrained by the tradeoffs of their ultimate use-case: compilation toolchains. Compilers solve a well-defined problem, and, thus, are deterministic and preserve semantics. For example, compilers typically have a limited lookahead and implement limited error-recovery, e.g. panic mode~\cite{dragonbook}, if any. 
In contrast, repair engines need to ``guess'' the meaning of an ill-formed program, as there may be multiple possible valid repair candidates (i.e., error-free variants) of a faulty program, but not all may match the user's intent. Furthermore, these repairs may constitute additions and deletions of user input, often faraway from the location where the error is detected.
Purely symbolic techniques often fail to distinguish between, or even generate, such viable candidate repairs. E.g., the state-of-the-art error-recovery tool \grmtools~\cite{ecoop20} non-deterministically picks from edit operation sequences within the same edit distance. It also focuses on the first error location found, and ignores repairs that involve edits preceding that error location. Our use-case requires generation of exhaustive repairs and a finer-grained ranking among multiple repairs.

\paragraph{\textbf{Neurosymbolic technique:}}
Increasingly, systems combine neural and symbolic approaches to leverage the advantages of both.
For example, Synchromesh~\cite{synchromesh} enforces symbolic constraints on a large pre-trained language model (LM) during decoding for natural language to code generation. These symbolic constraints remove common LM mistakes such as referencing unavailable names. Another effective approach is to use neural models that can guide otherwise symbolic approaches. For example, Kalyan et al~\cite{kalyan2018neural} showed that a model can help speed up search substantially by guiding branching decisions in deductive search.

\noindent\paragraph{\textbf{Neurosymbolic repair with \lamirage}} Based on the above observations, we design \lamirage, a
framework that can generate a last-mile repair engine tailored to a particular target
LC formula language using neurosymbolic techniques for proposed repair candidate enumeration and ranking. 
To motivate our design, we take inspiration from program synthesis~\cite{DBLP:journals/ftpl/GulwaniPS17}, where
the intent lies within the faulty program. The feasibility of program synthesis
rests on two pillars: (1)~controlling the search space through effective candidate enumeration
and (2)~ranking to pick
one from many candidates. To address these requirements, 
\lamirage has the following key properties:

\begin{itemize}
 \item {\emph{Effective enumeration.}} 
A na\"ive approach that
  enumerates and explores all possible options fails to scale because the
  search space is large. \lamirage exploits the fact that
  most errors are in certain \emph{unreliable language symbols}. This observation
  allows us to (1)~produce language-agnostic edit actions based on insertion/deletion/update
  of such unreliable symbols, and (2)~bias enumeration to these symbols, allowing 
  the engine to search deeper in the search
  tree, and, thus, examine repair candidates that involve fixes far from the
  reported error locations. 

\item {\emph{Complementary domain specific knowledge.}} The language-agnostic edit actions are syntax driven, editing the buggy formula to satisfy the grammar provided. However, non-context-free properties may also need to be enforced to produce a valid repair candidate. To enforce such properties,  \lamirage{} uses additional rules to 
generate repairs that eliminate certain classes of semantic errors, such as adjusting 
the number of parameters in a call to remove arity errors.

\item {\emph{Neural error localization and candidate selection.}}
In cases where the actual error location may lie outside the 
range considered by the symbolic engine,
\lamirage uses a \textit{pointer network}~\cite{vinyals2015pointer} to predict additional error location ranges. To enumerate candidates
at these locations, we consider a (location-specific) \emph{edit range} surrounding each predicted location. This symbolic relaxation allows us to take advantage of the predictions even if they are imperfect. The next challenge we address with a neural technique is
the selection of the the intended correct formula  among the set of generated repair candidates. Ranking only based on distance from the
  faulty formula is not sufficient, as this
  often leads to ties. We break such ties using \emph{neural selection}: we use
  a fine-tuned, pre-trained, deep-neural-network-based language model to guide
  us to the most natural formula. Our experiments
  (Section~\ref{sec:evaluation}) show the effectiveness of our neural localization and 
  selection approaches.

\end{itemize}

In contrast to existing symbolic state of the art, \lamirage{} offers more expressive repairs. In particular, while being
syntax-guided, \lamirage{} supports repairs to some semantic errors by allowing language developers to
easily integrate domain-specific strategies for identifying errors beyond those captured by a CFG. Additionally, 
\lamirage{} supports backtracking (carried out both symbolically and by a neural localizer) to perform
edits required for errors that have a root cause at a different location from where they are detected. Both of
these contributions allow \lamirage{} to substantially outperform the symbolic state of the art in our
evaluation.

From a neural perspective, \lamirage{} follows  prior work in the area of program
synthesis and program repair that use machine learning to improve their search process. The novelty of \lamirage{}
lies in its application of this combination of techniques to the problem of fixing last-mile errors in 
the low-code domain, paired with the specific setup (and insights) required to make neural models,
such as our localizer and ranker, work well in combination with an effective symbolic approach.

In summary, we make the following contributions:
\begin{itemize}
	
    \item We define the problem of \emph{last-mile repair}
    for buggy programs (\Cref{sec:problem}). We present
    a tractable formulation that approximates the target execution engine
    with a grammar and a set of constraints. We motivate the application of last-mile repair in the low-code domain (\Cref{sec:example}),
    where the grammar and constraints can successfully model the 
    target engine.

    \item We present \lamirage, a neurosymbolic approach that combines
    the strengths of both symbolic techniques (effective enumeration) and deep
    learning (natural ranking and long-range error localization) 
    to solve the last-mile repair problem in LC formula languages. 
    We show that more expressive repairs, paired with 
    effective use of neural models to complement
    the symbolic procedures, can improve the number of repairs in 
    our low-code evaluation.
    
    \item We developed concrete grammar and constraint approximations that are
    empirically effective for the domains of Excel and PowerFx. We evaluate \lamirage on two benchmark sets of 200 faulty formulas from each domain collected from help forums and
    system telemetry collected from the PowerApps and Excel teams at Microsoft. We compare \lamirage to the state-of-the-art neural and symbolic systems~\cite{BIFI, codex, codex-edit-api, ecoop20}, 
    and a commercially available alternative.
    In Excel, \lamirage's top repair candidate matched the ground truth
    repair in 174 out of 200 formulas, compared to 147 for the next best system, \codexedit. In PowerFx, \lamirage's top repair candidate matched the ground truth in 170 out of 200 formulas, compared to 106 for \codexedit. We also motivate our design with ablation studies.

    \item We release our benchmarks, gathered from real user forum posts and telemetry collected from our industrial partners, along with manually annotated ground-truth repairs.\footnote{The link will be made available once we have completed our privacy review.}

\end{itemize}

The paper is structured as follows. \Cref{sec:example} provides an example of
last-mile repair in a PowerFx formula. \Cref{sec:problem} presents the formal
problem definition and \Cref{sec:approach} details our approach.
\Cref{sec:evaluation} presents our experimental results and
\Cref{sec:use-and-design} discusses the choices language developers
can make in \lamirage{}. \Cref{sec:related} summarizes related work
and \Cref{sec:conc} provides closing remarks.

\section{Motivating Examples and Overview}
\label{sec:example}

\begin{figure*}[t]
\resizebox{\textwidth}{!}{
\begin{tabular}{lll}
\toprule
    & \textbf{Faulty expression at various stages of repair} 
    & \textbf{Description}
    \\
\midrule
P1: & \texttt{If (!IsBlank(LunchSeminar, UpdateContext(LunchSeminarVar \hlpink{:} LunchSeminar)))}
& Error location \hlpink{$:$}

\\ 
P2: & 
\texttt{If (!IsBlank(LunchSeminar\hlcyan{)}, UpdateContext(LunchSeminarVar \hlpink{:} LunchSeminar)))}
    & ADD \hlcyan{$)$}
\\ 
P3: & 
\texttt{If (!IsBlank(LunchSeminar\hlcyan{)}, UpdateContext(LunchSeminarVar\hlcyan{.}LunchSeminar\hlcyan{(})))}
   & REPLACE \hlpink{$:$} by \hlcyan{$.$} and ADD \hlcyan{$($}
\\ 
P4: & 
\texttt{If (!IsBlank(LunchSeminar\hlcyan{)}, UpdateContext(\hlcyan{\{}LunchSeminarVar : LunchSeminar\hlcyan{\}}))}
& ADD \hlcyan{$\{$} and REPLACE \hlpink{$)$} by \hlcyan{$\}$}\\ 
\bottomrule
\vspace{-8mm}
\end{tabular}
}
\caption{
\small
The user authors PowerFx formula $P1$, but the compiler reports an error at \hlpink{$:$}. 
To automatically fix this,  \lamirage first inserts a \hlcyan{$)$} to get $P2$, 
inducing the correct arity for the 
$\mathtt{IsBlank}$ function call. \lamirage then generates multiple repairs for the remaining 
errors, including $P3$ and $P4$, both of which have edit distance $3$ from $P1$. The
correct expression, $P4$, is ranked higher based on its naturalness by fine-tuned \codebert.
This example is a real user's PowerFx formula adapted for ease of exposition.
}
\label{fig:motivating}
\vspace{-2mm}
\end{figure*}

Popular low-code (LC) platforms, such as Excel and Power Apps, often expose
subsets of their functionality via drop-down menus in a graphical user interface.
However, to accomplish many tasks users must write formulas in the underlying LC
formula language. This can represent a significant challenge, as LC platform users
have varying degrees of programming experience and are typically attracted to such a platform
in part due to the relatively lower barrier to entry.  In particular, LC 
formula authors often struggle when identifying and manually fixing small errors in their formulas.
In this 
\emph{last mile} setting, the formula is
almost correct, but still requires a few tweaks to be fully correct. To illustrate
this common experience we will walk through the details of a Power Apps example,
where users write PowerFx formulas.

\begin{example}\label{ex:one}
Consider an incorrect PowerFx formula $P1$ (Figure~\ref{fig:motivating}).
This example is a real user's PowerFx formula adapted for ease of exposition.
The user is trying to update the context by assigning to
$\mathtt{LunchSeminarVar}$ the value of $\mathtt{LunchSeminar}$ if the latter
is not empty.
The formula $P1$ has errors: the compiler points to the colon
{`$:$'} as the culprit location. However, the root cause of the error is not at that
location.\qed
\end{example}

\looseness-1
Novices often struggle to fix such buggy formulas. This can result from multiple
factors, including: lack of user experience writing formulas, large formulas that compose multiple functions,  
lack of LC editor support for simple features such as syntax highlighting, ambiguous or complex
error messages from the underlying LC language toolchain. This combination of factors
can make spotting and fixing even small bugs a daunting task.

Furthermore, not all errors in formulas are due to simple typographical mistakes.
In $P1$, for example, one of the errors arose because the user was
not aware that a key-value pair needs to be enclosed within curly braces `$\{$'
and `$\}$'. Additionally, errors in LC formulas can also be semantic mistakes.
For
example, if we wrap the key-value pair in $P1$ with curly braces, 
the compiler will no longer report a syntax error, but type errors (and arity errors) remain: the
$\mathtt{If}$ function requires two or three arguments of appropriate types.

\begin{example} \label{ex:two}
The incorrect PowerFx formula $P1$ (Figure~\ref{fig:motivating}) is an
instance of the last-mile repair problem: the correct
formula ($P4$) is $3$ edits away from $P1$ and can be obtained by:

(1)~inserting a parenthesis `$)$' after $\mathtt{LunchSeminar}$,
(2)~inserting an opening brace `$\{$' before the key-value pair,
and
(3)~replacing the closing parenthesis `$)$' after the key-value pair by a closing brace `$\}$'.\qed
\end{example}

We now describe our framework, which allows us to generate such repairs automatically.

\looseness-1
\paragraph{\lamirage Overview} 

\begin{figure}[t]
    \includegraphics[width=0.9\columnwidth]{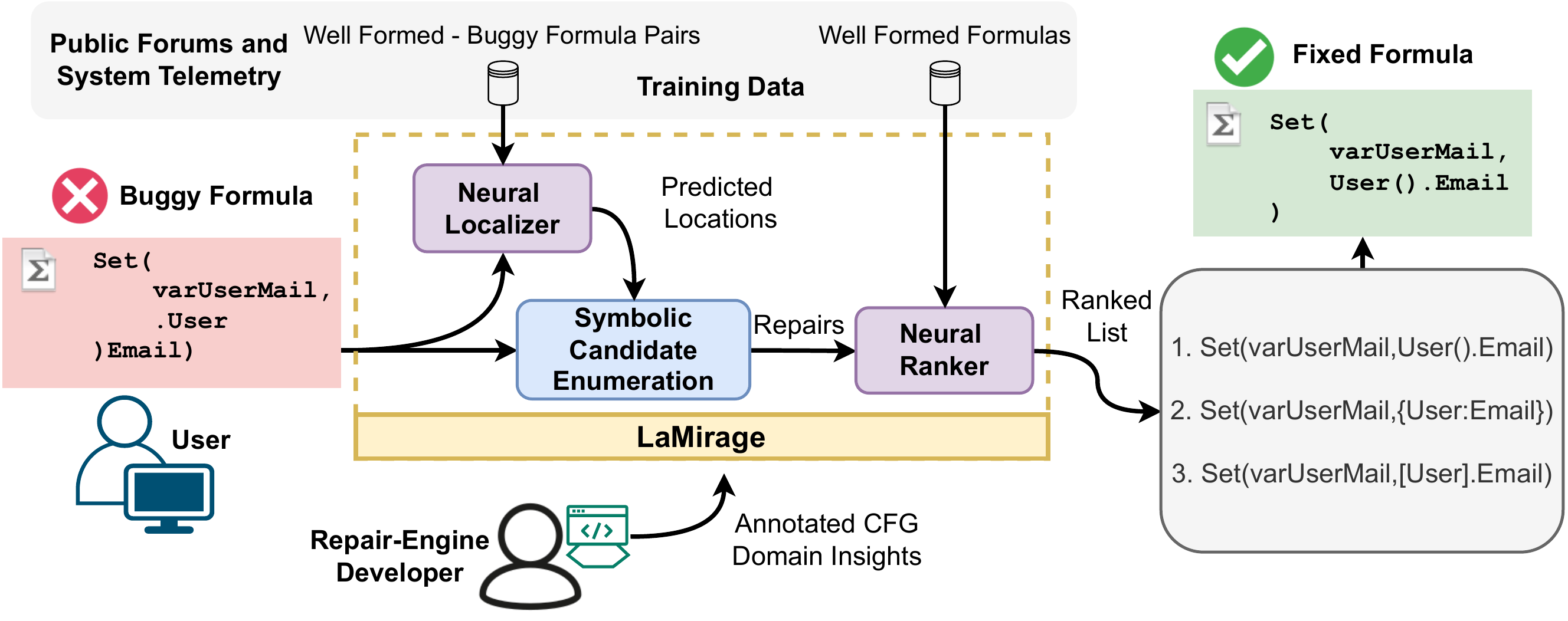}
\caption{
\lamirage uses three key components: a neural error localizer, a symbolic candidate enumerator, and a neural ranker. The neural localizer predicts error locations that augment deterministic backtracking to  perform edits far from the location where the error is identified -- for efficiency, we only use predict locations when our deterministic backtracking does not yield viable candidates. The symbolic enumerator produces candidate repairs that are guaranteed to satisfy the grammar and
constraints provided by the repair engine developer. Finally, a neural ranker,
trained on well-formed formulas from the domain, re-ranks candidate repairs. We train both neural models on formulas collected from public forums and industry partners' telemetry.
}
\vspace{-4mm}
\label{fig:architecture}
\end{figure}

\lamirage is a last-mile repair-en\-gine generator. A repair-engine generator
is a meta procedure that takes an annotated grammar approximating a target LC language and optional language-specific transformation rules and checks (which we view as constraints), 
and generates a repair engine that can
fix last-mile errors in formulas in the target LC language.
Figure~\ref{fig:architecture} presents the architecture of \lamirage. Developers
create a new repair engine by providing: (I1)~an annotated grammar that
specifies the \emph{unreliable} terminals in the specific LC language, 
(I2)~domain-specific insights, such
as repair rules, (I3)~a collection of paired well-formed and buggy formulas to train a neural error localizer,
and (I4)~a collection of well-formed formulas to fine tune
a ranker (\codebert\cite{FengCodeBertEMNLP2020}). Given a faulty formula, 
the engine predicts error locations using the neural error localizer. The
locations complement those identified symbolically. The engine then
enumerates repair candidates, ranked by the token edit distance from the
faulty formula. These candidates are obtained by inserting and/or deleting tokens
corresponding to \emph{unreliable} terminals in the faulty formula at the
error locations identified. Ties between candidate repairs are
resolved using fine-tuned \codebert to select the most \emph{natural}
candidate to return to the user.

The repair engines created by \lamirage are syntax-directed, working as modified LL parsers,
but the repairs produced are not limited to fixing 
syntactic errors but also common semantic errors.
A \lamirage-generated repair engine works like a
normal LL parser with the following modifications: \\

\indent - If the parser reaches a failure state, then instead of stopping, the repair
engine backtracks and attempts to insert or delete the ``unreliable'' tokens
defined in I1. These edit operations can fix many syntactic errors, and can also induce
semantic corrections (e.g. fixing function arity mistakes by re-associating call arguments using a
different parenthesization). \\

\indent - Each time the parser's internal state is updated, an external call to a
(appropriate) repair rule, defined in I2, is made, which can optionally
change the parser's state. These calls allow us to implement transformations that can induce
more complex semantic corrections that require domain knowledge about the underlying LC formula language
beyond that captured in the grammar annotations provided in I1. For example, these rules can
perform basic type casts in frequently misused function calls
or correct naming errors in the formula.

\begin{example}
	
Continuing from Example~\ref{ex:two}, we describe how we repair $P1$. The
\emph{unreliable} terminals for PowerFx correspond to punctuation tokens
(parentheses, curly braces, brackets, dots, commas, and colons) in our PowerFx 
implementation. During parsing $P1$, when the first comma is processed, a
domain-specific rule is triggered that enforces the
arity for the $\mathtt{IsBlank}$ call to
be $1$ by inserting a closing parenthesis. Thus, $P1$ is turned into $P2$.
Arity analysis is one domain-specific insight incorporated for the PowerFx
domain (I2).

Other examples include rules that fix misspelled function or variable names.

\looseness-1
The repair engine follows the steps of a regular parser until it hits the token
colon `$:$'. At this point, the parser backtracks to the point where the $d$-th
last reliable token was consumed. Here, $d$ is a parameter whose value lies in
$[1$--$4]$. Since $\mathtt{LunchSeminarVar}$ and $\mathtt{UpdateContext}$
(identifiers) are reliable, if $d=2$, we backtrack to the point 
after
$\mathtt{UpdateContext}$ is consumed. Our engine now enumerates several repairs that add
and/or remove punctuation tokens. While in this example, symbolic backtracking yielded
repair candidates, there may be formulas where the true error location lies in an earlier part of the formula outside the sybmolic backtracking depth. In such cases, \lamirage
makes use of neural error localization (described in \Cref{sec:neural-error-loc}) to predict error locations.

Following this methodology of backtracking and candidate enumeration, we generate
candidates $P3$ and $P4$. Since $P3$ and $P4$  are both at a token-edit distance
$3$ from $P1$, we break the tie using fine-tuned \codebert to find $P4$ as the
most natural repair.

Conceptually, this syntax-driven approach to enumerating
candidate repairs and then ranking the resulting repaired formulas based on a score (in this case, the edit distance and neural ranker score) has parallels with traditional program synthesis.
\qed

\end{example}

This example demonstrates a key insight behind \lamirage: successfully repairing
last-mile errors in LC formulas benefits from a combination of symbolic and neural techniques.
Without symbolic candidate enumeration, there is no \emph{guarantee} that the 
repair candidates produced are valid programs in PowerFx, without the neural re-ranking
of candidate repairs, the correct result $P4$ would not be identified as the most natural repair,
and without neural error localization,  locations far from the error state would not be considered
during formula editing.

\section{Last-mile Repair Problem}
\label{sec:problem}

Let $T$ be a target engine---e.g., a compiler, an interpreter, or a run-time
engine---that accepts or rejects programs. Informally, given a string $s$,
representing an ill-formed program, we seek to transform (fix) $s$ to another
nearby string $\sfixed$ that is accepted by $T$. Several values of $\sfixed$ may
exist, but we want the one that the user most likely intended.

The first challenge to fixing $s$ is to generate candidate strings that are
near to the original buggy input and will also be accepted by $T$. In most real-world scenarios, it is impractical or expensive to query $T$
repeatedly to identify valid fix candidates. Furthermore, we need
to have a practical way to produce these candidates.
Rather than consider all strings, we take inspiration 
from syntax guided synthesis and use a context free grammar $G$ 
as a constructive approximation of $T$. 

A context-free grammar
is a quadruple $G := (V, \Sigma, R, S)$, where $V$ is the the set of
non-terminals, $\Sigma$ is the set of terminals, $R \subset V \times (V\cup
\Sigma)^*$ is the set of production rules, and $S\in V$ is a special start
symbol.

A string $s \in \Sigma^*$ is accepted by $G$, or in the language defined by
$G$, denoted by $s \in L(G)$, if there exists a \emph{derivation} of $s$ in
$G$. A derivation of $s$ is a sequence of strings $S \rightarrow s_1
\rightarrow \ldots \rightarrow s_k \rightarrow s$, where each $s_i \in (V \cup
\Sigma)^*$, and for each $s_i \rightarrow s_j$, $s_j$ is obtained by replacing
a non-terminal $X$ in $s_i$ with $X_1, \ldots, X_n$ where each $X_i \in (V \cup
\Sigma)$ and $(X \rightarrow X_1\ldots X_n) \in R$.

Let $L_\mtext{full}$ be the language of all inputs
accepted by $T$. 
A smaller $L(G)$ implies a more efficient enumeration of candidate repairs
(as the space is smaller) but the reduction in overlap with $L_\mtext{full}$,
may place some candidate repairs out-of-scope. A larger $L(G)$ may increase
the scope of fixes, but may also result in spurious candidates outside of $L_\mtext{full}$. Given a fixed size of $L(G)$, we want to maximize
overlap with $L_\mtext{full}$, but one might allow spurious candidates
if it makes writing $G$ easier. Note that we can
apply $T$ on the final candidate repairs (after search) to filter out invalid candidates.

The second challenge is that $G$, by definition, can only cover context-free
properties of $L_\mtext{full}$. To address this challenge, we
introduce a set of context-sensitive constraints
$\mathcal{C}$ that are satisfied by all programs in $L_\mtext{full}$. 
A constraint
$C \in \mathcal{C}$ is a mapping: $\Sigma^* \mapsto \mathbb{B}$. Constraints
capture requirements that (1)~well-formed programs must satisfy, and (2)~are
not captured by $G$. Some examples are: (a)~programs should be type correct,
(b)~each variable should be defined, and (c)~every function or
operator name should be supported by $T$. Together,  
$G$ and the constraints $\mathcal{C}$ serve as a proxy for $T$.

Let $\mathcal{C_\mtext{full}}$ be the set of all constraints enforced by $T$.
Like $G$, $\mathcal{C}$ is an approximation. However, unlike $G$, 
$\mathcal{C}$ in practice is typically a sound over-approximation, meaning: $\forall s \in \Sigma^{*}. \mathcal{C}_\mtext{full}(s) \implies \mathcal{C}(s) \land \neg\mathcal{C}(s) \implies \neg\mathcal{C}_\mtext{full}(s)$. Evaluating $\mathcal{C}$
is more efficient than checking acceptance with $T$ (by construction), so 
we can use it during our candidate repair search. We present
more discussion around the tradeoffs in choice of $G$ and $\mathcal{C}$
in \Cref{sec:use-and-design}.

The third challenge is to model the user intent, i.e., quantifying the
likelihood of a candidate fix being the one that the user intends. We want to
maximize the probability $Pr(\sfixed \mid s)$ which quantifies the probability
that $\sfixed$ is the user-intended program given they wrote $s$. Our
assumption is that the user usually writes a program that is ``close'' to
what they intend. To quantify ``closeness'', we can use any distance metric
$\mathtt{dist}$ on strings. In this work, we use \emph{token edit distance} and
require a distance threshold $\delta$ that specifies the required closeness.
Among the set of programs that are within a distance of $\delta$ from $s$, we assume that
$Pr(\sfixed\mid s)$ is proportional to the prior probability, namely $Pr(\sfixed)$, of observing
$\sfixed$. Intuitively, we want to find a string $\sfixed$ as a repair that is ``close''
(according to distance function $\mathtt{dist}$ and the distance threshold
$\delta$) to the buggy program $s$, while making sure that it is ``valid''
(validated using the grammar $G$ and the constraints $\mathcal{C}$). In case of
multiple such candidates, we break ties by leveraging the probability distribution of ``natural''
programs that real users compose in the target language, defined by $Pr$.

Now we can formalize the problem statement as follows.
Given a grammar $G :=
(V,\Sigma,R,S)$, constraints $\mathcal{C}: \Sigma^*\mapsto\mathbb{B}$, a
distance measure $\mathtt{dist}: \Sigma^*\times\Sigma^*\mapsto\mathbb{R}^+$,
a distance threshold $\delta$,
and a string $s\in \Sigma^*$,
the {\em{last-mile repair problem}} seeks to find a string
$\sfixed\in\Sigma^*$ such that
$\sfixed\in L(G)$ and $\mathcal{C}(\sfixed)$, $\dist(\sfixed, s) \le \delta$,
and $\sfixed = \argmax_{X \in \{x \text{ s.t. } \mathtt{dist}(x, s) \le
    \delta\}} Pr(X)$, where $Pr$ is the probability distribution over
    human-composed strings in $L_\mtext{full}$.
\section{Repair-engine Generator}
\label{sec:approach}
We present the \lamirage framework, our specific solution to the
last-mile repair problem, and describe how
it can be instantiated to generate repair engines
for different LC formula languages.

The class of possible repairs can be large, and consequently, the search space
of potential repairs can be enormous. Therefore, we need to focus on classes
that represent a large set of common mistakes that low-code users make
when authoring programs in the target language.
This target-specific information is captured using the concept of {\em{unreliable
terminals}} and {\em{domain-specific parser state transforms}}. While unreliable terminals
capture pure syntax errors user's may make, domain-specific parser state transformers allow \lamirage to incorporate more semantic fixes.

Let $G := (V, \Sigma, R, S)$ be the grammar. \lamirage further assumes access to the following:
\begin{itemize}
    \item 
        A set $U\subset \Sigma$ of {\em{unreliable terminals}}: 
        A subset of the terminals is classified as unreliable based on their likelihood of being erroneously omitted or included in user-authored buggy formulas.
        For example, in formula languages, parentheses and/or punctuation marks are observed to be unreliable as user's often tend to misplace them in expressions. 
    \item 
        A set of {\em{domain-specific parser state transformations}}, where each transformation takes a parser state and returns a set of (modified) parser states. We will later see examples.
\end{itemize}

The subset $U$ and
domain-specific transformations 
are specified by the developer using \lamirage to instantiate a repair engine for their LC formula language.

Intuitively, \lamirage can be seen as a syntax-guided repair engine, generating repair
candidates by performing rule-based transformations (enumerating candidate valid
programs) on top of an LL parser that accepts strings in $G$.

The main differences from a regular LL parser are the following:

\begin{enumerate}

\item rather than produce a single parse, the repair engine explores
and produces multiple parses;

\item the repair engine backtracks when a failure state is reached
by the underlying LL parser;

\item the repair engine thereafter proceeds searching over valid candidates by
transforming unreliable terminals (and trusting reliable terminals to constrain
the search space);

\item at every step, the repair engine also calls parser state
transformers to get new parser states that can accumulate context sensitive information 
(e.g. number of arguments in current call); and

\item the repair engine limits the number of repairs it considers 
by tracking the edits (cost) it has already made (incurred) on the
input string. 
\end{enumerate}

Next, we formalize this intuition of an error-correcting parser.

\subsection{Parser States}

We describe the error-correcting parser using inference rules. The inference
rules operate on parser states. Given a grammar $G := (V, \Sigma, R, S)$, a
{\em{parser state}} is a 4-tuple $\tuple{A, T, p, c}$, where $A$ is the parsing
stack, $T$ is the stream of remaining tokens that need to be processed, $p$ is
the parse-tree constructed so far, and $c$ is the cost of the state. The
parsing stack $A$ is represented as a list, with the first element of the list
corresponding to the top of the stack. Similarly, $T$ is also represented as a
list, with the first element being the next immediate token. For convenience,
we use $\mtext{Stack}(x)$, $\mtext{RemTokens}(x)$, $\mtext{ParseTree}(x)$, and
$\mtext{Cost}(x)$ to refer to the parsing stack, remaining tokens, the
parse-tree, and the cost of a search-state $x$ respectively.

Let $G := (V, \Sigma, R, S)$ and the input string be $s$. Let $\mathtt{toks}$
denote the tokenization of $s$ represented as a list of tokens. The initial
state of the error-correcting parser is $([S,\$], \mathtt{toks}, p_0, 0)$,
where $\$$ is the end-of-sequence symbol, $p_0$ is parse tree containing just a
single (root) node $S$, and cost is $0$. Starting from this initial state, the
inference rules describe how states are updated. In some cases, multiple rules
may be applicable, or the same rule may result in multiple states. In these
cases, the interpretation is that of a {\em{non-deterministic choice}} - the
actual implementation considers {\em{all possibilities}} and explores all
states, providing a completeness guarantee with respect to $G$. The top-level algorithm, shown as
Algorithm~\ref{alg:overall-algorithm}, is just a specific strategy for applying
these inference rules. The goal is to start with the initial state and reach
the special state, $\mathit{accept}$, that is the terminating state for the
algorithm. Any state of the form $([],[],p,c)$, for any parse tree $p$ and cost
$c$, rewrites to the $\mathit{accept}$ state
(Figure~\ref{fig:state-transition-inference-rules}, detailed in \Cref{sec:repair-algorithm}).

Informally, starting from 
$([S,\$],\mathtt{toks},p_0,0)$, where $\mathtt{toks}$ is a 
tokenization of input string $s$, if we reach
$([],[],p,c)$, then $p$ will be a parse of some string
$\sfixed$ obtained by repairing $s$, and $c$ will be 
$\mathtt{dist}(\sfixed,s)$ (the cost of the repair).

\begin{figure}[t]
\small{
\begin{align*}
    \mtext{First}(t) &= \{t\} &\text{ if t is a terminal}\\
    \mtext{First}(X) &= \mtext{First}(\gamma_1) \cup \ldots \cup \mtext{First}(\gamma_n) &\text{ if $\forall i: X\rightarrow\gamma_i\in R$}\\
    \mtext{First}(s\gamma) &= \mtext{First}(s) \cup (\mtext{First}(\gamma) 
    \text{ if s can derive } \epsilon \text{ else }
 \emptyset)
    \\
    \mtext{Follow}(X) &= \mtext{First}(\delta)\ \cup
                              (\mtext{Follow}(Y) \text{ if } \delta \text{ can derive } \epsilon \text{ else } \emptyset)
                              & \text{ for all } Y \rightarrow \gamma X \delta \text{  rules in the grammar}
\end{align*}
}
    \vspace{-7mm}
	\caption{\small Fixed-point equations for the standard $\mtext{First}$ and
$\mtext{Follow}$ helper functions used in an LL parser for grammar $G :=
(V,\Sigma,R,S)$.}

\label{fig:first-follow-standard}
\end{figure}

\begin{figure}[t]
    \small{
    \begin{adjustbox}{width=1.\columnwidth}
    \begin{mathpar}

    \inferrule* [lab=T-Terminal-Match]
    { a = t \land a \in \Sigma \cup \{\$\}}
    {\tuple{a:A, t:T, p, c} \rightarrow \tuple{A, T, p, c}}

    \qquad\qquad
    \inferrule*[lab=T-Accept]
    { }
    {\tuple{[], [], p, c} \rightarrow \mtext{accept}}

    \\
    \inferrule* [lab=T-Non-Terminal-Expansion]
    { (a \in V) \land (a \rightarrow X_1 \cldots X_k) \in R \\
      t \in \mtext{First}(X_1\cldots X_k) \lor (t \in \mtext{Follow}(a) \land \epsilon \in \mtext{First}(X_1\cldots X_k)) \\
      p^\prime \text{is obtained from p by adding } X_1\cldots X_k \text{ as children of the leftmost leaf } a \text{ in p}}
    {\tuple{a:A, t:T, p, c} \rightarrow \tuple{[X_1, \cldots, X_k] {\texttt{++}} A, t:T, p^\prime, c}}
    \end{mathpar}
    \end{adjustbox}
    }
    \vspace{-6mm}
	\caption{\small Transition rules for the syntax-guided 
	repair engine states given a grammar $G :=
(V, \Sigma, R, S)$. Definitions for $\mtext{First}$ and $\mtext{Follow}$ are
provided in Figure \ref{fig:first-follow-standard}. The end-of-sequence token
is denoted by $\$$. Symbols $:$ and \texttt{++} corresponds to Haskell-style
list prepend and append syntax.}
    \vspace{-3mm}
	\label{fig:state-transition-inference-rules}
\end{figure}

\subsection{Repair Algorithm}\label{sec:repair-algorithm}

Algorithm \ref{alg:overall-algorithm} describes the overall approach. At a high level, 
we explore the search space using the
transition rules in Figure \ref{fig:state-transition-inference-rules} and
maintains a priority queue of states, ordered by their costs (Line
\ref{line-alg:init-priority-queue-overall-alg}).

The priority queue is initialized to contain just the initial state (Line
\ref{line-alg:initial-state-overall-alg}). Every time it encounters a state
corresponding to an accept state, it translates the parse-tree into a repair
and returns it to the user (Lines
\ref{line-alg:accept-start-overall-alg}-\ref{line-alg:accept-end-overall-alg}). Otherwise, it applies domain-specific strategies to obtain a set of new states in Line \ref{line-alg:domain-specific-overall-alg}. Domain-specific strategies are explained later in Section \ref{sec:domain-specific-parser-state-transformers}. The next step is to compute the set of next states using the transition rules from Figure \ref{fig:state-transition-inference-rules} (Line \ref{line-alg:error-1-overall-alg}). There are two conditions for the state
to be considered an error state (Line \ref{line-alg:error-2-overall-alg}). The set of next states ($N$) can be empty,  in which case no progress can be made without making edits, or the state
is part of the set of states predicted as possible error states by our neural error localizer (detailed later in Section~\ref{sec:neural-error-loc}).
If the state is considered to be an error state,  the algorithm then goes
into \emph{repair mode}. It uses a sub-procedure \mtext{EnumerateRepairs}
(detailed in Section~\ref{sec:core-enum}) to apply a correction on the error state and put
it back in the search queue (Line \ref{line-alg:repair-overall-alg}).

The transition rules in Figure~\ref{fig:state-transition-inference-rules}
describe a standard LL parser. The transition~\textsc{\small T-Terminal-Match}
handles the case when the top of stack is a token that matches the next token
in the input stream. The transition~\textsc{\small T-Non-Terminal-Expansion}
replaces a nonterminal $X$ at the top of stack by the list of elements from the
right-hand side of some production rule based on the lookahead. The standard
helper functions, $\mathit{Follow}$ and $\mathit{First}$, used in this rule are
obtained using fix-point computation over the equations shown in
Figure~\ref{fig:first-follow-standard}. Note that these two functions are pre-computed for a grammar.

\begin{algorithm}
\small
    \caption{Overall Repair Engine Algorithm. Given the list of tokens corresponding to a buggy string $s$, and a grammar $G$, it returns a list of repairs. The \$ symbol corresponds to the end-of-stream token. $\mtext{GenRepair}(s)$ converts the parse tree of state $s$ back into a string. The optional $S_\mtext{pred}$ correspond to states associated with error locations predicted by our neural error localizer -- which are only used when deterministic backtracking does not yield viable repair candidates.
    }
    \label{alg:overall-algorithm}
    \begin{algorithmic}[1]
 		\Procedure{\mtext{Repair}\,}{toks, G, $S_\mtext{pred} = \emptyset$}
 		\State $p_0$ \SymAssign{} \text{tree with a single node for }\mtext{StartSymbol}(G)
 		\State $s_0$ \SymAssign{} $\tuple{[\mtext{StartSymbol}(G) \$], \text{toks}, p_0, 0}$\Comment{{\small Initial State}} \label{line-alg:initial-state-overall-alg}
 		\State $P$ \SymAssign{} An empty priority queue \label{line-alg:init-priority-queue-overall-alg}
 		\State Insert $s_0$ into $P$ with cost 0.
 
 		\While{$P$ is not empty}
 		    \State Pop $s$ from $P$ with cost $c$
 		    \If{$s \rightarrow \mtext{accept}$}\label{line-alg:accept-start-overall-alg}
 		    \Comment{{\small Rules in Figure \ref{fig:state-transition-inference-rules}}}
 		        \State \textbf{yield} $\mtext{GenRepair}(s)$\label{line-alg:accept-end-overall-alg}
 		    \EndIf
 		    
 		    \State $S_D$ \SymAssign{} $\mtext{ApplyDomainStateTransformers}(s)$\label{line-alg:domain-specific-overall-alg}
 		    \State $N$ \SymAssign{} $\{s^{\prime\prime}\ |\ s^\prime \rightarrow s^{\prime\prime} \land s^\prime \in S_D\}$\Comment{{\small Rules in Figure \ref{fig:state-transition-inference-rules}}} \label{line-alg:error-1-overall-alg}
 		    
 		    \If{$N$ is empty or $s \in S_\mtext{pred}$}\label{line-alg:error-2-overall-alg}
 		        \State N \SymAssign{} $\mtext{EnumerateRepairs}(s)$\label{line-alg:repair-overall-alg}
 		    \EndIf 

 		    \ForEach{$s^\prime$ in $N$}
 	            \State Insert $s^{\prime}$ into $P$ with cost $\mtext{Cost}(s^{\prime})$
 		    \EndFor
 		\EndWhile
 		\EndProcedure
 	\end{algorithmic}
\end{algorithm}

\begin{figure}
    \centering
    \includegraphics[width=0.9\columnwidth]{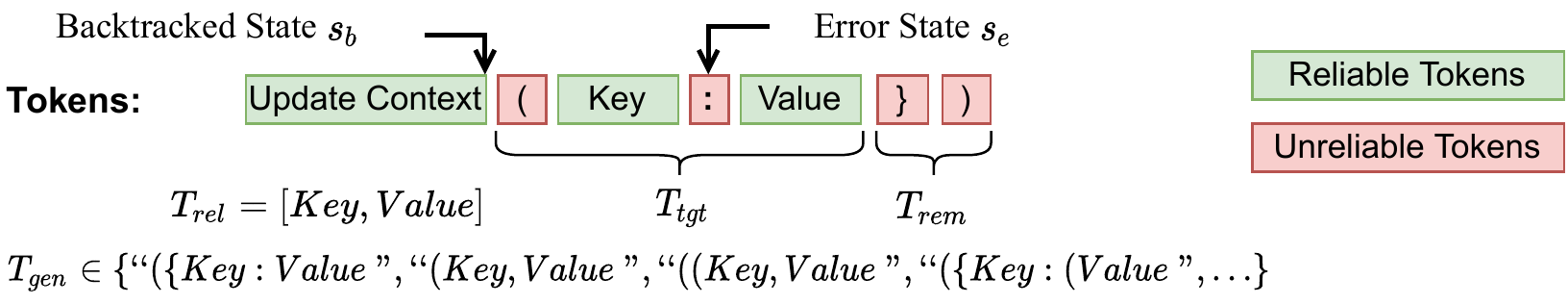}
    \caption{Illustration of repair enumeration, along with token
    reliability and deterministic backtracking.} 
    \label{fig:enumeration-diagram}
\end{figure}

We extend this initial rule set with those in
Figure~\ref{fig:state-transition-enumerate-repairs} to convert the LL parser
into a syntax-guided repair candidate enumerator, which \lamirage uses to 
produce repair candidates (which then need to be ranked).
Algorithm~\ref{alg:repair-candidate-enumeration} presents a strategy for
applying these new extension rules. 

\subsection{Repair Candidate Enumeration}\label{sec:candidate-enum}

Algorithm \ref{alg:overall-algorithm} calls \mtext{EnumerateRepairs} to
\emph{repair} an error state, as described in  Algorithm 
\ref{alg:repair-candidate-enumeration}.

We use $\tau(s) = \tuple{s_0, \ldots, s}$ to refer to the \emph{trace} of $s$ i.e. to the sequence of states resulting from repeated application of the transition rules in Figure \ref{fig:state-transition-inference-rules} starting from the initial state $s_0$, as detailed in Algorithm \ref{alg:overall-algorithm}, and ending at $s$.  Note that the trace is just a history of parser states that we assume the algorithm saves.
We say a token $t$ is {\em{unreliable}} if \texttt{t.value == a} for some unreliable terminal $a$; i.e., when the value attribute of the token corresponds to an unreliable terminal.

Given an erroneous search state $s_e$, we first identify
the portion of input processed so far that can be edited to induce 
candidate repairs. \lamirage accomplishes this using 
deterministic backtracking.

\subsubsection{Deterministic Backtracking}\label{sec:deterministic-backtracking}
To produce the editable formula range, \lamirage deterministically
\emph{backtracks} to an ancestral state $s_b \in \tau(s_e)$. The state $s_b$
corresponds to the state in $\tau(s_e)$ where the $d^\mtext{th}$ previous
\emph{reliable token} was added to the parse-tree of $s_e$, or if no such state
exists, $s_b$ is the first state in $\tau(s_e)$. Consider the example in Figure
\ref{fig:enumeration-diagram}. The error state corresponds to the point after
\texttt{Key} is consumed. With $d=2$, the backtracked state corresponds to the
point right after \texttt{UpdateContext} was consumed. The constant $d$ here
refers to the \emph{backtracking depth}, which
controls the extent to which we can reach and edit previously processed tokens.

\looseness-1
We use $T_\mtext{tgt}$, $T_\mtext{rem}$ and $T_\mtext{rel}$ to refer to three
special lists of tokens. Let $t_\mtext{r}$ be the first reliable token in
$\mtext{Tokens}(s_e)$. $T_\mtext{tgt}$ is the list of tokens in
$\mtext{Tokens}(s_b)$ up until and including $t_\mtext{r}$ (note that
$\mtext{Tokens}(s_e)$ is guaranteed to be a suffix of $\mtext{Tokens}(s_b)$).
$T_\mtext{rem}$ is the list of tokens after, and excluding $t_\mtext{r}$ in
$\mtext{Tokens}(s_e)$. Finally, $T_\mtext{rel}$ is the list of all reliable
tokens, in order, in $T_\mtext{tgt}$. This logic is encapsulated within the
$\mtext{Backtrack}$ procedure in Line
\ref{line-alg:backtrack-call-enum-repair} of \Cref{alg:repair-candidate-enumeration}. Consider the example in
Figure~\ref{fig:enumeration-diagram}. $T_\mtext{tgt}$ corresponds to the four
tokens \texttt{(}, \texttt{Key}, \texttt{:}, and \texttt{Value}.
$T_\mtext{rem}$ is the list of tokens following \texttt{Value}. $T_\mtext{rel}$
is the list of two reliable tokens in $T_\mtext{tgt}$ - \texttt{Key} and
\texttt{Value}. 

Tokens in $T_\mtext{tgt}$ that are unreliable (i.e. not in $T_\mtext{rel})$ are candidates
for editing during candidate enumeration. Meanwhile, tokens outside of
$T_\mtext{tgt}$, such as tokens in $T_\mtext{rem}$, remain untouched.

Scoping edits to a subset of tokens in $T_\mtext{tgt}$, in contrast to considering
the entire program prefix, helps constrain the search space for candidate repairs,
allowing \lamirage to quickly produce viable repairs. While deterministic
backtracking is efficient, it is not complete: there may be required edits outside of the  range identified. \lamirage uses neural techniques to address this challenge.

\subsubsection{Neural Error Localization}\label{sec:neural-error-loc}
\begin{figure}[t]
\includegraphics[width=0.8\columnwidth]{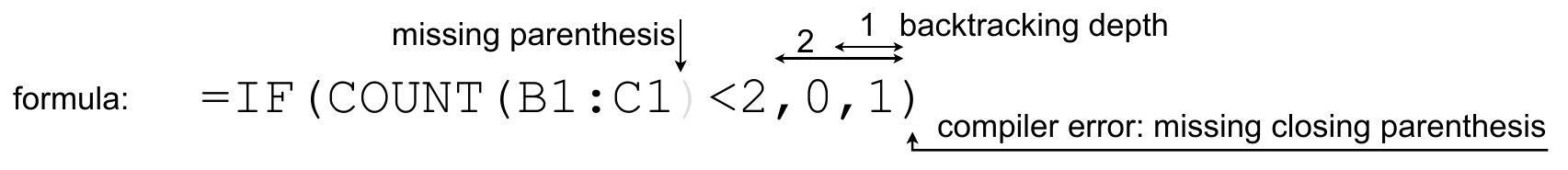}
\caption{\small 
The true location of an error may be relatively far  from where
the compiler detects an error. For example, in this formula the missing parenthesis
is not identified until the end of the formula, as \texttt{COUNT} is variadic.
\lamirage{} uses a pointer-network to predict such error locations.
}
\vspace{-2mm}
\label{fig:techinque:nel}
\end{figure}

Compilers often raise errors at locations far away from the actual source of mistake~\cite{traver2010compiler}.
In Figure~\ref{fig:techinque:nel}, the user forgot a closing parenthesis for the COUNT function
in their Excel formula. However, the Excel compiler will parse
the formula until the end-of-stream token is encountered, and will \emph{then} raise a missing parenthesis error. Why is no error reported earlier? The function \textsc{COUNT} is variadic, so the compiler greedily accepts the string  contents as valid arguments for the function call. A purely symbolic repair technique might rely on backtracking to identify possible repair locations, however, as the formula grows, so does the backtracking depth required. This increase in depth can substantially increase the search space. To mitigate this problem, we leverage neural methods to complement our deterministic backtracking.

We train a Pointer Network~\cite{vinyals2015pointer} to predict error locations in arbitrary length formulas by learning distributions over input tokens, which has been shown in previous work~\cite{vasic2018neural} to be effective for error localization in general purpose programming languages. Specifically, we take a corpus of well-formed formulas in the corresponding language and then generate broken variants by introducing synthetic errors. These errors are introduced by randomly adding, deleting, or changing unreliable tokens in the formula. We then train the pointer network to predict the locations (token indices) where these edits were performed as a function of the broken formula. At prediction time, we take the top 5 locations predicted by the network, though in practice the network mostly predicts 1 or 2 locations. For each predicted location, we take the parsing state up to that location when processing the input and treat it as an error state $s_e$. We can then apply our deterministic backtracking strategy starting from each such state. As a result the pointer network need not be perfect to still provide useful enumeration locations, in contrast to prior work~\cite{vasic2018neural} that jointly localizes and repairs. The symbolic candidate enumerator can then perform edits as before to yield new candidates. 

To mitigate the increase in the size of the search space, as a result of
additional candidate edit locations, \lamirage{} employs a \emph{fall-back} strategy.
Specifically, \lamirage first attempts to repair a buggy program using only the deterministic backtracking strategy, and if no viable repair candidates are produced, \lamirage then employs the neural error localizer to predict error states. These neurally predicted error states augment the set identified by the deterministic approach and are passed transparently to the candidate enumeration algorithm $\mtext{EnumerateRepairs}$.

This use of the neural localizer mirrors previous lines of work
in program synthesis and program repair that 
integrate model-based approaches to restrict or prioritize elements in their
search space, for example ~\cite{long2017automatic, yu2019learning}. The novelty in our approach is in showing a simple pointer-network-based localizer can work well in the low-code domain, particularly when employed in a fall-back strategy.

\subsubsection{Candidate Enumeration with Guarantees}\label{sec:core-enum}
$\mtext{EnumerateRepairs}$ returns all states $s_r$ that are the same as $s_b$
except for their remaining token sequences, where $T_\mtext{tgt}$ is replaced
by a new sequence of tokens $T_\mtext{gen}$ (Line
\ref{line-alg:repair-state-enum-repair}). $T_\mtext{gen}$ satisfies the
constraint that it can be obtained from
$T_\mtext{tgt}$ by inserting and/or deleting unreliable tokens. Additionally, the
parsing stacks of the repaired states $s_r$ are guaranteed to be able to derive
the string corresponding to $T_\mtext{gen}$ as a prefix. Furthermore, the modified states
are guaranteed to satisfy custom checks (as they are applied prior to returning candidates).
These guarantees are a key benefit of \lamirage compared to purely neural alternatives.
Figure~\ref{fig:enumeration-diagram} shows a few possibilities for $T_\mtext{gen}$
that are generated by the algorithm.

\begin{figure}[t]
    \small{
    \begin{adjustbox}{width=1.\columnwidth}
    \begin{mathpar}
    \inferrule* [lab=T-Unreliable-Terminal]
    { a \in \Sigma \land a \in U}
    {\tuple{a:A, T_\mtext{gen}, T_\mtext{rel}} \rightarrow \tuple{A, T_\mtext{gen}\texttt{++}[a], T_\mtext{rel}}}
    \qquad
    \inferrule* [lab=T-Reliable-Terminal]
    { a \in \Sigma \land a \notin U \land a = t_\mtext{rel}}
    {\tuple{a:A, T_\mtext{gen}, t_\mtext{rel}:T_\mtext{rel}} \rightarrow \tuple{A, T_\mtext{gen}\texttt{++}[a], T_\mtext{rel}}}
    \\
    \inferrule* [lab=T-Non-Terminal]
    { (a \in V) \land (a \rightarrow X_1 \cldots X_k) \in R \\
      (t_\mtext{rel} \in \mtext{FirstReliable}(X_1\cldots X_k)\ \lor
      (t_\mtext{rel} \in \mtext{FollowReliable}(a) \land \epsilon \in \mtext{FirstReliable}(X_1\cldots X_k)))}
    {\tuple{a:A, T_\mtext{gen}, t_\mtext{rel}:T_\mtext{rel}} \rightarrow \tuple{[X_1, \cldots, X_k] {\texttt{++}} A, T_\mtext{gen}, t_\mtext{rel}:T_\mtext{rel}}}
    \\
    \inferrule*[lab=T-Accept]
    { }
    {\tuple{a:A, T_\mtext{gen}, []} \rightarrow \mtext{accept}}
    \end{mathpar}
    \end{adjustbox}
    }
	\vspace{-4mm}
    \caption{\small State transition rules for $\mtext{EnumerateRepairs}$ given a
    grammar $G := (V, \Sigma, R, S)$ and unreliable terminals $U \subset
    \Sigma$. Definitions for $\mtext{FirstReliable}$ and
    $\mtext{FollowReliable}$ are provided in Figure
    \ref{fig:first-follow-reliable}. Symbols $:$ and \texttt{++} correspond to
    Haskell-style list prepend and append syntax.}
    \label{fig:state-transition-enumerate-repairs}
	\vspace{-6mm}
\end{figure}

\looseness-1
At the core of $\mtext{EnumerateRepairs}$ is the ability to enumerate
valid $T_\mtext{gen}$. This is achieved by repeated application of the
transition rules in Figure \ref{fig:state-transition-enumerate-repairs}. The
first rule states that if the top of the stack is an unreliable
terminal, add it to the generated token sequence so far. The second rule
covers the case when the top of the stack is reliable; in this case it must match against the next expected reliable token as per $T_\mtext{rel}$. The
third rule governs the production rules chosen to expand the top of the stack
when it is a non-terminal. This is analogous to the non-terminal rule in the
standard parsing transition rules in Figure
\ref{fig:state-transition-inference-rules}. The only difference is the use of
$\mtext{FirstReliable}$ and $\mtext{FollowReliable}$. These two functions are similar to their standard counterparts $\mtext{First}$ and
$\mtext{Follow}$ used in Figure \ref{fig:state-transition-inference-rules}. The
only difference is that they are only concerned with reliable terminals, not
all terminals.

However, recall that, as per the last-mile repair definition, we do not want just any $T_\mtext{gen}$ that is possible. It must be
within some edit-distance of the original. Thus we ensure that the edit-distance (Lines
\ref{line-alg:edit-distance-exact-1-enum-repair}-\ref{line-alg:edit-distance-exact-2-enum-repair}), or its lower-bound estimate (Lines
\ref{line-alg:edit-distance-lb-1-enum-repair}-\ref{line-alg:edit-distance-lb-2-enum-repair}) is less than the hyper-parameters $\mtext{MaxGlobalCost}$ and
$\mtext{MaxLocalCost}$, which restrict the maximum edit distance.

\begin{figure}[t]
\small
\begin{algorithm}[H]
    \caption{Enumerating repairs given an error state $s_e$}
    \label{alg:repair-candidate-enumeration}
    \begin{algorithmic}[1]
 		\Procedure{\mtext{EnumerateRepairs}\,}{$s_e$}
 		\State $s_b, T_\mtext{rel}, T_\mtext{tgt}, T_\mtext{rem}$ \SymAssign{} $\mtext{Backtrack}(s_e)$ \label{line-alg:backtrack-call-enum-repair}
 		\State $P$ \SymAssign{} An empty queue
 		\State Insert $\tuple{\mtext{Stack}(s_b), [], T_\mtext{rel}}$ into $P$.
 		\State \mtext{repairs} \SymAssign{} empty list
 		
 		\While{$P$ is not empty}
 		    \State Pop $\tuple{A, T_\mtext{gen}, T_\mtext{rel}}$ from $P$
 		    \If{$\tuple{A, T_\mtext{gen}, T_\mtext{rel}} \rightarrow \mtext{accept}$}\Comment{Figure \ref{fig:state-transition-enumerate-repairs}}
 		        \State $c_\mtext{r}$ \SymAssign{} $\mtext{EditDist}(T_\mtext{gen}, T_\mtext{tgt}) + \mtext{Cost}(s_b)$ \label{line-alg:edit-distance-exact-1-enum-repair}
 		        \State $s_\mtext{r}$ \SymAssign{} $\tuple{\mtext{Stack}(s_b), T_\mtext{gen}\texttt{++}T_\mtext{rem}, \mtext{ParseTree}(s_b), c_\mtext{r}}$\label{line-alg:repair-state-enum-repair}
 		        \If{$c_\mtext{r}$ $\leq$ $\mtext{MaxGlobalCost}$}\label{line-alg:edit-distance-exact-2-enum-repair}
 		            \State Add $s_\mtext{r}$ to \mtext{repairs}
 		        \EndIf
 		        \State \textbf{continue}
 		    \EndIf
 		    \State \mtext{NextStates} \SymAssign{} $\{x\ |\ \tuple{A, T_\mtext{gen}, T_\mtext{rel}} \rightarrow x\}$ \Comment{Figure \ref{fig:state-transition-enumerate-repairs}}
 		    \ForEach{$\tuple{A^\prime, T_\mtext{gen}^\prime, T_\mtext{rel}^\prime} \in \mtext{NextStates}$}
 		        \Comment{Compute lower-bound on edit-distance}
 		        \State $c$ \SymAssign{} $\{\mtext{EditDist}(T_\mtext{gen}^\prime, p)\ |\ p \text{ is a prefix of } T_\mtext{tgt}\}$ \label{line-alg:edit-distance-lb-1-enum-repair}
 		        \If{$c$ $\leq$ $\mtext{MaxLocalCost}$}\label{line-alg:edit-distance-lb-2-enum-repair}
 		            \State Append $\tuple{A^\prime, T_\mtext{gen}^\prime, T_\mtext{rel}^\prime}$ to $P$ 
 		        \EndIf
 		    \EndFor
 		\EndWhile
 		
 		\State \Return \mtext{repairs}
 		\EndProcedure
 	\end{algorithmic}
\end{algorithm}
\vspace{-8mm}
\end{figure}

\begin{figure}[t]
\small{
\begin{align*}
    \mtext{FirstReliable}(t) &= \{t\} \text{ if t is a terminal and t is reliable}\\
    \mtext{FirstReliable}(t) &= \emptyset \text{ if t is a terminal and t is unreliable}\\
    \mtext{FirstReliable}(X) &= \mtext{FirstReliable}(\gamma_1) \cup \ldots \cup \mtext{FirstReliable}(\gamma_n)
                             \text{ where } X \rightarrow \gamma_1 \ldots X \rightarrow \gamma_n \text{ are production rules}\\
    \mtext{FirstReliable}(s\gamma) &= \mtext{FirstReliable}(s) \cup (\mtext{FirstReliable}(\gamma)
                                   \text{ if s can derive } \epsilon \text{ else }\emptyset)
    \\
    \mtext{FollowReliable}(X) &= \mtext{FirstReliable}(\delta)\ \cup 
                               (\mtext{FollowReliable}(Y) \text{ if } \delta \text{ can derive } \epsilon \text{ else } \emptyset)
                  \\ & \qquad            \text{ for all } Y \rightarrow \gamma X \delta \text{ rules in the grammar}
\end{align*}
}
\vspace{-6mm}
\caption{\small Fixed-point equations for $\mtext{FirstReliable}$ \& $\mtext{FollowReliable}$.}
\vspace{-6mm}
\label{fig:first-follow-reliable}
\end{figure}

\subsection{Domain-Specific Parser State Transformers}\label{sec:domain-specific-parser-state-transformers}

The repairs generated by Algorithm \ref{alg:repair-candidate-enumeration} 
are guaranteed to satisfy $G$. However, valid formulas in the target language
must also satisfy constraints $\mathcal{C}$, which capture semantic properties, such as 
correct typing or using only defined variable names. Fixing formulas that do not satisfy
these constraints often requires context-sensitive information and additional
knowledge about the underlying LC formula language beyond that reflected in $G$.
A key innovation of \lamirage{}, in contrast to state-of-the-art symbolic repair
systems like \grmtools{}, is the increase in repair expressiveness -- which can capture some
semantic errors -- enabled by our use of $\mathcal{C}$.

We introduce the concept of \emph{domain-specific repair strategies} via \emph{parser state transformers} to tackle these classes of errors, as captured by the call to
$\mtext{ApplyDomainStateTransformers}$ in Line \ref{line-alg:domain-specific-overall-alg} of
Algorithm~\ref{alg:overall-algorithm}. Transformers are essentially a collection of symbolic transition rules, that create new parser states from a given state, or flag a state as an \emph{error state}, denoted as $\bot$. Given the input parser state, the
$\mtext{ApplyDomainStateTransformers}$ function simply returns the set of states obtained by applying all the eligible transition rules, or an empty set if the input state is flagged as an error state by \emph{any} of the rules. To support different use cases, \lamirage{} supports transformers that are called at different points in the input processing procedure. For example, error state transformers are called when an error state is raised, other
transformers may trigger on a particular token/rule.

The transition rules for various strategies that can be implemented in our
framework are listed in Figure \ref{fig:domain-specific-strategies-rules}. All
the rules are based on the grammar fragment described in Figure
\ref{fig:relevant-grammar-fragment}. This fragment captures function calls and
basic expressions as allowed by the LC domains of Excel and PowerFx. Next, we
explain each of our strategies individually.

\begin{figure}[t]
\small{
    \begin{align*}
        \text{FuncCall} &::= \text{FuncName}\ (\ \text{ArgsList}\ ) 
        & \text{FuncName} &::= \text{ident}\\
        \text{ArgsList} &::= \epsilon\ |\ \text{Arg}\ \text{ArgsListTail}
        & \text{ArgsListTail} &::= \epsilon\ |\ ,\ \text{Arg}\ \text{ArgsListTail}\\
        \text{Arg} &::= \text{Expr}
        & \text{Expr} &::= \text{Var}\ |\ \text{Constant}\ |\ \text{BinaryExpr}\ |\ \ldots\\
        \text{Var} &::= \text{ident} & &
    \end{align*}
}
\vspace{-6mm}
    \caption{\small Relevant fragment of grammar used for Excel/PowerFx corresponding to function calls and expressions.}
    \label{fig:relevant-grammar-fragment}
\vspace{-2mm}
\end{figure}

\begin{figure}[t]
    \resizebox{\columnwidth}{!}{
    \small{
    \begin{adjustbox}{width=.95\columnwidth}
    \begin{mathpar}
    \inferrule* [lab=T-Arity]
    { a \in \{\text{ArgsList}, \text{ArgsListTail}\}\\
      f = \mtext{CurFunc}(p)\\n = \mtext{CurNumArgs}(p, f)\\
      (n \geq \mtext{MaxArity}(f) \land t.kind \notin \mtext{Follow}(a) \lor\\
       n < \mtext{MinArity}(f) \land t.kind \in \mtext{Follow}(a))
    }
    {\tuple{a:A, t:T, p, c} \rightarrow \bot}
    \\
    \inferrule* [lab=T-Combine-Tokens]
    { \text{Tokens } t_1 \text{ and } t_2 \text{ are eligible to be combined into } t_3 \\
      \text{with cost } c^\prime}
    {\tuple{A, t_1:t_2:T, p, c} \rightarrow \tuple{A, t_3:T, p, c + c^\prime}}
    \\
    \inferrule* [lab=T-Symbol]
    { a \in \{\text{FuncName, Var}\} \\
      t.val \notin \mtext{AvailableSymbols} \\
      \tuple{t^\prime, c^\prime} \in \mtext{AvailableCorrections}(t.val)}
    { \tuple{a:A, t:T, p, c} \rightarrow \tuple{a:A, t^\prime:T, p, c + c^\prime} }
        \\
    \inferrule* [lab=T-Symbol-Fail]
    { a \in \{\text{FuncName, Var}\} \\
      t.val \notin \mtext{AvailableSymbols} \\
      \mtext{AvailableCorrections}(t.val) = \emptyset}
    { \tuple{a:A, t:T, p, c} \rightarrow \bot}
    \\
    \inferrule* [lab=T-Typing]
    { p^t = \mtext{ComputeTypes}(p) \\
     p^t \text{ has typing errors}\\
     \tuple{p^\prime, c^\prime} \in \mtext{ComputeTypeRepairs}(p^t)}
    { \tuple{A, T, p, c} \rightarrow \tuple{A, T, p^\prime, c + c^\prime}}
    \\
    \inferrule* [lab=T-Typing-Fail]
    { p^t = \mtext{ComputeTypes}(p) \\
     p^t \text{ has typing errors}\\
     \mtext{ComputeTypeRepairs}(p^t) = \emptyset}
    { \tuple{A, T, p, c} \rightarrow \bot}
    \end{mathpar}
    \end{adjustbox}
    }
    }
	\vspace{-6mm}
    \caption{\small Inference rules for various domain-specific, context-sensitive repairs based on the grammar
    fragment in Figure
    \ref{fig:relevant-grammar-fragment}. The symbol $\bot$ denotes an error state.
    }
    \label{fig:domain-specific-strategies-rules}
	\vspace{-6mm}
\end{figure}

\subsubsection*{Arity Analysis} Most formulas in
languages like PowerFx/Excel use built-in functions, which have a fixed minimum
and maximum arity. The \texttt{IsBlank} function in the motivating example in
Figure \ref{fig:motivating} has a minimum and maximum arity of 1, and thus
repairing P1 involves inserting a parenthesis after the first argument to
\texttt{IsBlank} which is \texttt{LunchSeminar}.

The rule \textsc{T-Arity} in Figure \ref{fig:domain-specific-strategies-rules} captures the arity analysis strategy that allows for such repairs. Essentially, given an input parser state, it first computes the current unclosed function $f$, and the number of arguments parsed so far for $f$, denoted by $n$, by analyzing the parse-tree $p$ using the convenience functions \mtext{CurFunc} and \mtext{CurNumArgs} respectively. Then, it flags the input parser state as an error state if the top of the stack is either ArgsList or ArgsListTail (Figure \ref{fig:relevant-grammar-fragment}) and one of two cases hold: (1) $n$ $\geq$ $\mtext{MaxArity}(f)$, and the next token is going to force the parse of another argument, and (2) $n < \mtext{MinArity}(f)$, and the next token does not indicate the start of a new argument. Whether or not a new argument is going to be parsed can be checked by checking the membership of the kind of the next token against the \mtext{Follow} set of the non-terminal at the top of the stack, as described in the rule \textsc{T-Non-Terminal-Expansion} in Figure \ref{fig:state-transition-inference-rules}.

How does this help in repair? Since the input parser state is flagged as an error state, this 
triggers \mtext{EnumerateRepairs} in Algorithm \ref{alg:overall-algorithm} in Line \ref{line-alg:repair-overall-alg}. Thus, this strategy will be able to fix arity errors by inserting/deleting unreliable tokens (punctuation).

\cheapsubsectionit{Combining Tokens.} An\-other common class of errors in both our domains involves incorrect tokenization due to presence of extra whitespace.
For example, in Excel, if a space is included between the $<$ and $=$ symbols, the whole
string is tokenized into two separate tokens, instead of the more likely $<=$
(less-than-equal) token, and will raise a syntax error (reported by the compiler as ``missing operand''). In both our domains, these binary operators are reliable terminals/tokens; hence Alg.~\ref{alg:repair-candidate-enumeration} can't generate repairs for such errors on its own and requires a domain specific strategy.

The strategy is formalized as the rule \textsc{T-Combine-Tokens} in Figure
\ref{fig:domain-specific-strategies-rules}. Essentially it says that
if the next two tokens are eligible to be combined into a new token with cost $c^\prime$, return a
new state where the two are combined and $c^\prime$ is added to the cost. Thus, this strategy directly modifies the remaining token stream. In our implementation,
we use this rule to combine tokens corresponding to relational operators such
as $<=$, $==$, and $>=$.

\cheapsubsectionit{Fixing Symbol Errors.}
This strategy helps generate repairs for errors where function names or variables are misspelled (\texttt{\small IsBlnk} instead of \texttt{\small IsBlank}), or synonyms for functions are used, such as the use of \texttt{\small Length} instead of \texttt{\small Len} in the second example in \Cref{fig:examples}.

The strategy is formalized as the rule \texttt{T-Symbol} in Figure \ref{fig:domain-specific-strategies-rules}. If the top of the stack corresponds to the non-terminals FuncName or Var in the grammar in Figure \ref{fig:relevant-grammar-fragment}, and the next token $t$'s value is not present in the set of available symbols, denoted by \mtext{AvailableSymbols}, and a corrected token $t^\prime$ is available with cost $c^\prime$, then a new state is returned where the next token is replaced with $t^\prime$.  The set of symbols \mtext{AvailableSymbols} can be determined from the runtime context within the GUI/IDE interface to the underlying domain. The function \mtext{AvailableCorrections} returns available symbols within a threshold edit-distance of the value of the original token $t$, which captures the misspelling case, and built-in functions that are known synonyms of the value of $t$. If no such correction is available, then the state is flagged as an error state (rule \textsc{T-Symbol-Fail}). Note that this will invoke \mtext{EnumerateRepairs} but none of its repairs would fix the issue and overall, no repairs would be returned by Algorithm \ref{alg:overall-algorithm}.

\cheapsubsectionit{Fixing Type Errors.}
The final strategy helps generate repairs for typing errors. Specifically, types are computed for the parse-tree of the input parse state, and if there is a type error, one of two things can happen: (1) if a repair is available in terms of a fixed parse-tree with cost $c^\prime$, then a new state is returned with the fixed parse tree and an additional cost of $c^\prime$, and (2) if a repair is not available, the state is flagged as an error state. The scenarios are captured by rules \textsc{\small T-Typing} and \textsc{\small T-Typing-Fail} in Figure \ref{fig:domain-specific-strategies-rules} respectively. Note that a repair on a parse-tree must apply to a node which is completely parsed i.e. it has no non-terminal leaf nodes. An example of a repair is explicit type conversion, such as converting an int to a string to enable concatenation with another string.

\cheapsubsectionit{Ease of Use.}
\lamirage{} enables repair-engine developers to create new domain-specific
strategies with a few lines of code -- this increase in repair
expressiveness beyond syntax fixes, despite being syntax-guided, is a key
contribution of \lamirage{}. For example, the equality operator in Excel (=) is
different from that in other languages such as Python and Java (==); confusing
these is a common mistake for some Excel users. To implement this DSS, a
developer need only define a few constants and conditions; \textit{Trigger
token}: "=", \textit{condition}: "Next token in the input stream is also =", and
the \textit{Transformation}: "skip the next = token from the input stream". Most
such DSS are reusable across domains with few changes. 

\subsection{Ranking Repairs by Naturalness}

Algorithm \ref{alg:overall-algorithm} is guaranteed to return the repairs in
 increasing order of cost, which is the edit-distance from the original
buggy program in our implementation. However, a scenario may arise when there
are multiple repairs within the allowable edit-distance, in which case we need to
select the most \emph{natural} repairs to show to the user. We fine-tune a
pre-trained language model, \codebert \cite{FengCodeBertEMNLP2020}, to
approximate the probability that a formula would be written by a user. \codebert is pre-trained on millions of aligned natural
language and code snippets across programming languages
such as Python, and Java, so we need to fine-tune it for our LC domains.

A simple way to fine-tune \codebert would be to train it with the
\emph{causal-LM} objective \cite{DaiCausalLMObjectiveNeurIPS2015} i.e. train it
to predict a formula one token at a time, by taking into account the tokens
generated so far. Then the product of the associated probabilities with every
generated token can be used for ranking the formula. Since the bulk of our
algorithm is focused towards producing repairs involving insertion/deletion of
unreliable tokens or punctuation, we can restrict the causal-LM objective to
only train the model to predict contiguous unreliable sequences given the list
of tokens before and after the target sequence. An  issue arises here: the
frequency distribution of punctuation tokens is highly skewed: e.g., a
parenthesis occurs more frequently than a curly brace. Thus training with this
objective over the available well-formed formulas introduces a bias in the
models towards more frequently occurring tokens and will not work
well, in our experience. To address this, we introduce a new setup
that mitigates this bias. Specifically, we break this task further into
predicting a single unreliable token given the prefix and suffix lists of
tokens, turning it into a classification task, where we can appropriately
balance the training dataset by undersampling and oversampling as necessary. To
rank repairs, we sum the negative log-probabilities of predicted tokens, and use
this to break edit-distance ties. In our experience, this modified setup worked
well for both domains and allowed us to use relatively modest amounts of data
during training.

Now that we have our full approach, \Cref{thm:result} summarizes the properties of our repairs.

\begin{theorem}\label{thm:result}
Let $G$ be the grammar provided to \lamirage and $\mathcal{C}$ be the set of domain specific constraints provided to complement $G$. Let $L_\mtext{full}$ be the language of inputs accepted by $T$, the target execution engine.
Jointly, $G$ and $\mathcal{C}$ approximate $L_\mtext{full}$, as described in \Cref{sec:problem}.
Let $\sfixed$ be a repair returned by \lamirage{}. 
Let $\mathtt{dist}: \Sigma^{*} \times \Sigma^{*} \mapsto \mathbb{R}^{+}$ 
be an edit distance metric between two strings. Let $\delta \in \mathbb{R}^{+}$ be the maximum permissible edit distance. Then $\sfixed \in L(G) \land \mathcal{C}(\sfixed) \land (\mathtt{dist}(s, \sfixed) \le \delta)$.
If a ranking model is used and has learned the appropriate prior distribution
over $L_\mtext{full}$, then $\sfixed$ maximizes $\mtext{Pr}(\sfixed)$.

\end{theorem}

\section{Evaluation} \label{sec:evaluation}
We present the results of our
empirical evaluation of an implementation of \lamirage
on two critical LC domains: Excel and PowerFx.
First, we describe our
experimental setup and methodology, including a description of our datasets and
the baselines we compare against. 
We also carry out a set of ablation studies (\Cref{sec:ablation}) to evaluate the impact of different design decisions on \lamirage.

\subsection{Benchmarks and Datasets}\label{sec:evaluation:methodology:datasets} 
We evaluate performance on two LC languages with significant user bases: Excel and PowerFx.

\cheapsubsection{Benchmarking.}
We created a benchmark set of 200 Excel formulas by gathering buggy formulas
listed in the publicly available third-party Excel forum MrExcel~\cite{mrexcel}.
We performed a similar collection of 100 PowerFx formulas from the official
PowerApps help forum~\cite{paforum}. We then added 100 PowerFx formulas
collected by the PowerApps team at Microsoft, who used basic system
telemetry to passively log anonymized PowerFx formulas written by real
anonymized users. The combined set of 200 PowerFx buggy formulas (and their
groundtruth solutions) constitute our PowerFx benchmark.  We manually annotated
the ground truth for all formulas.

To avoid introducing bias to the selection of formulas for our benchmarks, we adhered to the following procedure. For forum sourced benchmarks, we sampled uniformly at random from formulas scraped that did not pass the domain's parser/analyzer.
For telemetry benchmarks in the case of PowerFx, we sampled 500
formulas from our industrial partners' telemetry that did not pass PowerFx's analyzer -- we removed any broken formula that was incomplete (i.e. it was the result of a user still in the process of writing). For both forum and telemetry formulas, we removed formulas for which we could not 
unambiguously determine the ground truth solution manually.
The final collection of formulas include error such as: unmatched delimiters, invalid function call syntax (including
invalid spaces, extra commas, and supurious symbols),  invalid function call arity (both
excess and insufficient arguments), incorrect types, malformed references (such as sheet,
cell, and range references in Excel), malformed records (in PowerFx), invalid operator uses (including operators without operands or invalid use
of an operator in a function call), malformed relational operators (such as using
incorrect syntax for a comparison operation), and inappropriate string quoting.
As part of our contribution, we are releasing these benchmarks.

For our evaluation, we consider a candidate repair to be successful if it matches the ground-truth formula (after normalizing for white-space and casing, whenever not relevant).

\cheapsubsection{Training datasets.}
To produce training data for methods that require it, we relied on MrExcel forum posts and PowerApps forum posts.
We restrict ourselves to training data that is disjoint from the formulas used in our evaluation benchmarks.
We extracted formulas present in user posts. To improve the quality of PowerFx formulas collected, in
that domain we restrict ourselves to text in {\small{\texttt{<code>}}} HTML
tags. We perform semi-automated curation using manually written rules to remove
partial formulas or inputs outside of the
language domain targeted. We use a native parser for Excel and PowerFx,
respectively, to label formulas collected as parseable or un-parseable. 

From this collection, we prepared 
267,653 Excel formulas and 29,154 PowerFx formulas that satisfy the domain's analyzer and can be used
for language modeling and baseline training. In addition, we 
collected 27,501 Excel formulas and 1,183 PowerFx formulas that are rejected by the domain's analyzer
and are used during baseline training.

\subsection{Baselines}
We compare \lamirage to state-of-the-art symbolic and neural approaches. We produce up to 50 candidate repairs with each system. 
We report the number of benchmarks that are successfully repaired
if we consider the top 1, 3, and 5 candidate repairs produced. 
We base our cutoffs (up to 5 candidates) on working memory capacity~\cite{cowan2001metatheory} which has been applied for other recommendation systems~\cite{henley2018human}, though more studies are warranted.

\cheapsubsection{Our configuration.}
Our evaluation implementation of \lamirage includes
domain specific strategies for arity analysis and fixing symbol
errors for both Excel and PowerFx,
combining tokens strategy for Excel, and repairing ill-formed cell references in Excel. In terms of ranking, our implementation of \lamirage ranks candidate repairs lexicographically based on their edit-distance and language-model score. For neural error localization, we use the implementation as described in Section~\ref{sec:neural-error-loc}. We train the Excel and PowerFx neural error localizers on
pairs of original formula and synthetically broken formula, where we introduce at most 3 and 5 errors, respectively.

We set the local and global edit distances for \lamirage to 3 and a per-formula timeout
of 10 seconds, such that we only consider \lamirage candidates produced within
the timeout.

\cheapsubsection{Symbolic approaches.}
We consider a state-of-the-art symbolic error recovery system and the publicly available error recovery in Excel desktop. No such feature is available for PowerFx.

\paragraph*{\grmtools.} We use \grmtools, a parser framework that exposes the official implementation of a symbolic state-of-the-art parsing recovery technique~\cite{ecoop20}. \grmtools produces a set of one or more edit operations at one or more locations in the original input code. When there are multiple locations, \grmtools applies the first edit operation at location $i$ before generating the set of repairs at location $i + 1$. To produce a repaired candidate, we take the top edit operation at each location and apply it to the input formula. \grmtools does not rank repairs, as long as they have the same edit distance, and produces the edit operations in non-deterministic order for each location. For each formula, we run \grmtools 50 times and take the set of repairs returned in the given order. This approach is based on correspondence with the authors (via Github issues) for approaches to enumerate more than 1 repair candidate. To account for non-determinism, we repeat our \grmtools evaluation 10 times and report the best performance across cut-offs.

\paragraph*{\exceldesktop Error Recovery} We compare against the error recovery provided by Excel desktop Version 2203, which can correct errors such as adding missing closing parentheses. 

\cheapsubsection{Neural approaches.} We compare to three neural systems that represent two popular approaches in neural program
repair:  1) task-specific models and 2) large pre-trained language models.

\paragraph*{Break-It-Fix-It (\bifi).} \bifi iteratively trains an encoder-decoder based
neural code fixer and breaker -- the latter is used to improve the fixer that generates repair candidates. Both the fixer and breaker are implemented using transformers~\cite{vaswani2017attention}. We refer the interested reader to the associated paper~\cite{BIFI} for more details. We train \bifi on our Excel and PowerFx data and set the maximum generation length to 10 tokens beyond the input length.

\paragraph*{\codex} We use OpenAI's REST API to conduct experiments
with \codex~\cite{codex}, a state of the art large language model trained on code. 
We provide \codex with the following prompt:
\vspace*{-1mm}
\begin{lstlisting}[basicstyle=\ttfamily\scriptsize]
##### Fix bugs in the below code \n ### Buggy <domain> \n <buggy-code> \n ### Fixed <domain>
\end{lstlisting}
\vspace*{-1mm}
\noindent{where} we replace \texttt{<domain>} with Excel or PowerFx, and
the \texttt{<buggy-code>} with the formula we want to repair. To use \codex's few-shot learning abilities, we
include three manually written examples of buggy and fixed code from the
appropriate domain chosen to cover common mistakes.  We
rank \codex produced repairs based on their average log probability.

The three examples cover common mistakes: missing parentheses,
extra parentheses, and extra commas. These as they represent common user mistakes. Note that we also experimented with zero-shot learning
and found that including our predefined examples improved performance across the board.
Our prompt design did not exhaust the token limit but rather included
relevant examples. \citet{synchromesh} showed that similarity of examples with target task is important.

\paragraph*{\codexedit}  On March 22nd 2022, OpenAI released a new version of Codex designed for the task of editing existing user inputs~\cite{codex-edit-api}, rather than completing prompts.
 We use the REST edit API provided by OpenAI. In contrast to the traditional Codex API, the edit API does not take a standard prompt but rather takes an ``instruction'' parameter, which states (in natural language) what the model should do. We use the phrase \texttt{``Fix bugs in the <domain> code''} as the instruction and replace \texttt{``<domain>''} with Excel or PowerFx, as appropriate. Additionally, note that this API does not return multiple possible candidates but rather a single result. To mitigate this restriction, we call the edit API 50 times to produce up to 50 candidate repairs for each formula.

For \codex/\codexedit, we use temperatures 0, 0.1, 0.3, 0.4, 0.5, and 0.7. We report the best performance for each cutoff in our main evaluation; detailed breakouts in supplementary materials.

\subsection{Results}\label{sec:evaluation:e2e}

\begin{table*}[t]
\resizebox{\columnwidth}{!}{
\begin{tabular}{llrrrc|rrrc}
\multirow{2}{*}{\textbf{System}} & \multicolumn{1}{l}{\multirow{2}{*}{\textbf{Type}}} & \multicolumn{4}{c|}{\textbf{Excel}}                                                                                          & \multicolumn{4}{c}{\textbf{PowerFx}}                                                                                       \\ \cline{3-10} 
                                 & \multicolumn{1}{c}{}                                        & \multicolumn{1}{r}{\textbf{Top-1}} & \multicolumn{1}{r}{\textbf{Top-3}} & \multicolumn{1}{r}{\textbf{Top-5}} & \textbf{Time} & \multicolumn{1}{r}{\textbf{Top-1}} & \multicolumn{1}{r}{\textbf{Top-3}} & \multicolumn{1}{r}{\textbf{Top-5}} & \textbf{Time} \\ \hline
\exceldesktop                    
& 
Symbolic                                                    
& 83                                 & 83                                 & 83                                 
& 
-             
& 
\multicolumn{1}{r}{-}              & \multicolumn{1}{r}{-}              & \multicolumn{1}{r}{-}              
& 
-             
\\
\grmtools                         
& 
Symbolic                                                    
& 
97                                 & 104                                & 108                                
& 
427.6         
& 
98                                 & 110                                & 113                                
& 
351.4         
\\
\bifi                             
& 
Neural                                                      
& 
115                                & 130                                & 134                                
& 
363.1             
& 
34                                 & 45                                 & 48                                 
& 
592.8             
\\
\codex                            
& 
Neural                                                      
& 
111                                & 156                                & 160                                
& 
1651.8             
& 
86                                 & 117                                & 132                                
& 
1997.9             
\\
\codexedit                      
& 
Neural                                                      
& 
147                                & 163                                & 165                                
& 
5806.6             
& 
106                                & 137                                & 140                                
& 
6417.6             
\\
\textbf{\lamirage}                         
& 
NeuroSymbolic                                               
& 
\textbf{174}                       & \textbf{182}                       & \textbf{182}                       
& 
\textbf{32.1}             
& 
\textbf{170}                       & \textbf{177}                       & \textbf{177}                       
& 
 \textbf{134.4}            
\end{tabular}


}
\caption{\textit{More repaired formulas:} \lamirage can successfully repair more formulas than baseline approaches across all top-K cutoffs in both domains. 
For \codex{} and \codexedit{} 
we combine results across temperatures, reporting
the best value for each cutoff. The temperature with
best performance for top-1, top-3, and top-5 correspond to
0.3, 0.7, 0.7 (\codex in Excel);
0.4, 0.7, 0.7 (\codex in PowerFx);
0.1, 0.3, 0.4 (\codexedit in Excel);
and 0.0, 0.4, 0.5 (\codexedit in PowerFx);
We also report the median time (in milliseconds) to produce the repair candidates for each formula. \lamirage{} can produce its candidate repairs faster than other systems. 
}
\label{table:performance:e2e}
\vspace{-6mm}
\end{table*}

\Cref{table:performance:e2e} shows the number of formulas from our benchmark set that are successfully repaired by each system. 
In the Excel domain, we see that \exceldesktop, which only produces a single repair candidate for each faulty formula, performs worst, repairing only 83 of
the 200 formulas. The symbolic state-of-the-art \grmtools repairs  up to 108 formulas, but still lags the neural and neurosymbolic approaches. In particular, \grmtools fails to repair many formulas where the edit is far away from the location where the error state is raised. In terms of neural approaches, we find that \bifi, which is trained specifically for our task and domain, can repair substantially more formulas than the symbolic approaches (across all cutoffs).  \bifi also outperforms \codex at top-1, but repairs fewer formulas at more lenient cutoffs. \codexedit, which is designed for the task of editing user input, improves over both \bifi and \codex, outperforming both approaches over all top-K cutoffs. \lamirage, which combines the advantages of symbolic candidate enumeration with neural error localization and ranking, fixes the most formulas across all systems. \lamirage fixes 27  more formulas at the top-1 cutoff than the next best system, \codexedit.

In the PowerFx domain, 
\lamirage similarly outperforms all baselines across all cutoffs, repairing 64 more formulas at the top-1 cutoff compared to the next best system (\codexedit).
All systems with data-driven components, in this case neural approaches and \lamirage, experienced a drop in performance compared to the Excel domain, though \lamirage experienced a smaller drop. The purely symbolic approach \grmtools repaired a comparable number of formulas in both domains. We believe this is due to the fact that while PowerFx is a growing language, publicly available PowerFx code corpora is significant smaller than that for Excel. This challenge in data availability is reflected in our own training data as well. Neural approaches that benefit from significant pre-training (e.g. \codex and \codexedit) experienced a smaller drop than \bifi, which was trained specifically on the PowerFx data we collected. \lamirage, which combines symbolic and neural benefits, experienced the smallest drop in performance (4 fewer formulas repaired at top-1 cutoff) compared to the next best approach (\codexedit{}, which repaired 41 fewer at top-1 cutoff).

In both domains, we notice that \codex{} and \codexedit{} results improve substantially between top-1 and top-3 cutoffs (while still trailing \lamirage{}) and stabilizing thereafter. We believe these large pretrained models fail to account for differences in low-code formulas, as their training corpus primarily consist of programs written in general purpose languages. Both engines are capable of producing viable candidates, but are not suitably distinguishing between them.

We also compute
the median time to produce 
all candidate repairs for each formula in our benchmarks.
We exclude \exceldesktop as error recovery is exposed as a pop-up which requires user interaction, thus invalidating time measurement. \codex and \codexedit use the OpenAI
REST API, so they include network time, 
and we compute the minimum time across temperatures for each benchmark before summarizing
as a median. For \grmtools, we run the tool 50 times on each formula 
to obtain 50 candidate repairs, as such we add these times up. These repetitions incur
repeated processing within \grmtools{}, unavoidable without significant modifications to the tool. In both domains, the median time for \lamirage is substantially lower.

\subsection{Ablation Study Over \lamirage}\label{sec:ablation}
We now present results (\Cref{table:ablation:all}) that explore the impact of different design choices in \lamirage.

\begin{table*}[t]
\small
\begin{tabular}{llrrr|rrr}
\multicolumn{1}{c}{\multirow{2}{*}{\textbf{Ablation}}} & \multicolumn{1}{c}{\multirow{2}{*}{\textbf{System}}} & \multicolumn{3}{c|}{\textbf{Excel}}                                                                           & \multicolumn{3}{c}{\textbf{PowerFx}}                                                                       \\ \cline{3-8} 
\multicolumn{1}{c}{}                                   & \multicolumn{1}{c}{}                                 & \multicolumn{1}{r}{\textbf{Top-1}} & \multicolumn{1}{r}{\textbf{Top-3}} & \multicolumn{1}{r|}{\textbf{Top-5}} & \multicolumn{1}{r}{\textbf{Top-1}} & \multicolumn{1}{r}{\textbf{Top-3}} & \multicolumn{1}{r}{\textbf{Top-5}} \\ \hline
\multirow{3}{*}{Enumeration}                           
& 
\codebert 
& 
135                                & 137                                & 137                                 
& 
125                                & 130                                & 131                                \\
& 
\codex 
& 
128                                & 128                                & 128                                 
& 
112                                & 121                                & 121                                \\
& 
Whole-Prefix                                          
& 
71                                 & 97                                 & 110                                 
& 
\multicolumn{1}{r}{116}             & \multicolumn{1}{r}{141}             & \multicolumn{1}{r}{150}             \\ 
\hline
DSS                                                    
& 
No DSS                                               
& 
138                                & 145                                & 145                                 
& 
153                               &  164                                & 165                                \\ 
\hline
\multirow{2}{*}{Neural}                                
& 
No Neural Ranker                                     
& 
173.9                              & 182.2                             & 182.6                              
& 
150.9                             & 167.8                              & 169                                \\
& 
No Neural Localizer                                  
& 
167                                & 175                                & 175                                 
& 
159                                & 166                                & 166                                \\ 
\hline
Full                                                   
& 
\textbf{\lamirage}                                             
& 
\textbf{174}                                & \textbf{182}                                & \textbf{182}                                 
& 
\textbf{170}                                & \textbf{177}                                & \textbf{177}                               
\end{tabular}

\caption{\small 
\textit{\lamirage outperforms ablations:}
We consider the following ablations: (a) purely neural (\codebert{}, \codex{}) (b) naive symbolic
enumeration, (c) no domain specific strategies (DSS), (d) no neural ranker, and (e) no localizer.
In all but one case, \lamirage{} repairs more formulas across all top-K cutoffs.
We find that neural ranking (on average) does not play as big a role
in Excel as there are fewer candidates with
edit-distance ties and so neural ranking tie-breaks are less important.
}
\label{table:ablation:all}
\vspace{-6mm}
\end{table*}

\paragraph{\textbf{Enumeration ablation}}
We compare \lamirage 
to two ablations that replace \lamirage's candidate enumeration with
neural enumeration. \emph{\codebert} and \emph{\codex} in \Cref{table:ablation:all}
use \codebert  and \codex, respectively, to generate candidate repairs. 
To use these neural models for candidate enumeration,
when \lamirage{} reaches an error state, we make the corresponding neural model predict
the sequence of unreliable tokens bounded by the reliable tokens at the particular
error location (see Section~\ref{sec:candidate-enum} for details on this bounding).
In the case of Codex, we prompt the model to produce the sequence of unreliable tokens
as a completion to a prompt which corresponds to the formula prefix. Zero-shot prediction worked better for this task of generating unreliable token sequences.
In the case of \codebert, we predict the next unreliable token, given the prefix/suffix
of the formula, and we add a beam-search layer to output sequences (up to stop token).

We also compare \lamirage to an ablated version (\textsc{Whole-Prefix} in the table) that uses symbolic backtracking and candidate enumeration but considers the entire formula prefix (up to where the error state was identified) when generating candidate repairs. This is in contrast to \lamirage's approach of identifying edit ranges based on surrounding reliable tokens.

Our results show that \lamirage's enumeration outperforms all ablated versions across all top-K cutoffs in both domains. Ablation with \codebert outperformed the version that uses \codex for candidate enumeration in both domains as well. The ablated version with whole-prefix deterministic backtracking solves much fewer benchmarks as the search space explodes leading to time outs.

\paragraph{\textbf{Domain specific strategies}}
Next, we evaluate the impact of domain specific strategies (DSS) on \lamirage's
performance. Our results show
that removing DSS substantially decreases the number of benchmarks
solved in the Excel domain across all top-K cutoffs. DSS play a major role in the number of benchmarks solved in PowerFx across all cutoffs, 
but this benefit is smaller in this domain relative to Excel. 
DSS plays a bigger role in Excel owing to idiosyncrasies in its formula language that 
are difficult to capture with just a grammar approximation. For example, 
Excel's parser does not recognize a function call if there
is a space between the function name and the opening parentheses
-- a DSS, consisting of a single line of code, can resolve this. Relatedly, malformed cell references, another frequent user error, is more easily handled using DSS as compared to a grammar. Additional DSS can be added to \lamirage{} to further improve coverage of user scenarios.

\paragraph{\textbf{Neurosymbolic Ranking and Localization}}
Finally, we compared \lamirage to two ablations that remove
the neural error localizer and the neural ranker.
Without the ranker, the ablated version cannot distinguish between
candidate repairs with the same edit distance. So
we run the ablated version without neural ranker 100 times, compute results, and
report the average across each cutoff.

In Excel,
full \lamirage outperforms the version without neural localizer but is comparable
to the version without neural ranker. This behavior is
due to the fact that we generate fewer repair candidates with edit-distance ties  
in Excel, so there
is less opportunity to exploit neural ranking for tie breaking. 
In contrast, both neural
ranking and localization play a significant role in PowerFx.

\subsection{Discussion}
We discuss some takeaways around error analysis and neural approaches.

\cheapsubsection{Error Analysis.}
We consider the formulas in each domain that we cannot repair with \lamirage{} after considering the top-5 candidates. In Excel, we find that we can increase our coverage to 4 additional formulas if we change \lamirage hyperparameters: a deeper  backtracking would cover 3 formulas and a longer repair timeout would cover 1 more formula.
Extending our approximate grammar $G$ with new productions would cover 4 additional formulas.
Finally, if we add more type constraints to $\mathcal{C}$ and
token preprocessing DSS we would cover the remaining 9
formulas. 

In the case of PowerFx, increasing the set of type constraints and DSS would increase coverage to 14 additional formulas. If we extend our approximate grammar, we would cover 2 of the remaining formulas.
For one formula, our neural localizer
incorrectly predicts that a reliable token is the source of the error -- improving the localizer would resolve this issue. If we perform deeper deterministic backtracking, we 
can solve 3 more formulas. If we increase the edit distance threshold, we solve one more.
The remaining two formulas require inserting reliable tokens.

The solutions proposed here for increasing coverage all come with tradeoffs. By increasing backtracking depth, timeouts, and edit distance thresholds, expanding the approximate grammar, and adding more DSS and constraints, we can increase the number of formulas repaired at the expense of efficiency. \lamirage{} does not make a decision on these fronts, but rather lets the language developer choose the appropriate tradeoffs for their use case.

\cheapsubsection{Neural vs \lamirage.}
In our evaluation, \lamirage outperformed neural baselines
such as \codexedit. Cases where \codexedit failed (but \lamirage succeeded) provide interesting insights into challenges of purely neural approaches.

\begin{sloppypar}
Given the Excel formula \lstinline|=IF(B6="", "",|,
\codexedit (even at temperature=0.0) returns a degenerate candidate repair that nests the user's input repeatedly 
 \lstinline|=IF(B6="","",IF(B6="",...|. The fix simply requires a closing parenthesis. On other occasions,
 \codexedit adds spurious additional code. For example, given the formula
 \lstinline|=LEN(MID(A2,1,SEARCH("<",A2))|, \codexedit returns \lstinline|=LEN(MID(A2,1,SEARCH("<",A2)-1))| which changes
 the computation. This is especially challenging for LC settings, where the user is less likely to spot errors due to their lack of development experience. In other cases, \codexedit fails to recognize that there is an error
 altogether. For example, given \lstinline|=B2< =EDATE(TODAY(),-33)|, 
 \codexedit returns the original input rather than remove
 the space between the less than and equals operator (caught by \lamirage{}'s constraints $\mathcal{C}$).
 \end{sloppypar}
 
Training a neural model for low code language repair
can improve performance but may be challenging due to data availability. 
Additionally, fine-tuning large pretrained models like \codex
and \codexedit, while appealing, raises resource challenges (e.g. Codex has 12 billion parameters). Alternatively, developers can rely on paid APIs, like OpenAI's,
but this can represent substantial costs for high-volume low-code platforms (e.g. OpenAI's completion API costs 6 cents per 1k tokens). 
These models are also expensive at inference (prediction) time and unlikely to
to be deployed in resource-constrained environments for the 
foreseeable future.

Neural models will continue to improve, but symbolic systems can also be improved through
the addition of more domain knowledge. In practice, systems like \lamirage{}, which combine
both approaches, are a particularly interesting point in the design space. \lamirage{} can quickly (and cheaply, in terms of compute resources) produce effective repairs, but at the one-time cost of the language developer providing an annotated CFG and DSS.

\cheapsubsection{Applicability.}
We evaluated \lamirage{}'s applicability on two popular 
low-code languages: Excel and PowerFx. The former counts
hundreds of millions of daily users, the latter is the language used 
in PowerApps,
one of Microsoft's fastest growing offerings.
While the ideas presented in \lamirage{} may provide a good
starting point for tackling similar last-mile repair tasks
in other languages, such as small Python expressions, we leave that
for future work and keep
our focus on the low-code domain. It is worth highlighting that
the low-code domain is \emph{not} restricted to a single language,
but rather consists of different
platforms with different languages -- for example, Salesforce lightning, Creatio, Google sheets, Google AppSheet, Mendix, and other low-code platforms all have their own languages.

\section{Instantiating the Framework}
\label{sec:use-and-design}
\lamirage{} is a framework that can be instantiated
to create a specific repair engine for a particular
language. 
This instantiation is guided by the language developer,
and influences the kind of repairs that will be produced 
for consumption by the end-user.

By design, the framework supports many choices/components so that
a language developer can choose the tradeoffs that are right for their
domain and platform. To illustrate these choices, we will discuss the
\emph{necessary} components, along with additional extensions that
the language developer may choose to instantiate.

First, the developer needs to provide \lamirage{} with an annotated CFG. The CFG annotations correspond to marking terminals as unreliable, and potentially setting different
edit costs for different unreliable tokens. If the user does not specify per-token edit costs, 
we use a default value of 1. Most language developers
already have a CFG for their domain, or can craft one that
targets the subset of the language they believe
may benefit most from repairs. Similarly, in our experience,
language developers can quickly identify a set of unreliable tokens for their domain -- often those that are associated
with punctuation. For example, the Roslyn\footnote{a .NET compiler} team immediately noted that braces were unreliable tokens in the C\# domain.

After receiving the annotated CFG, \lamirage{} can produce a repair engine capable of fixing syntax errors (as captured by the CFG) by performing edits on unreliable tokens (as captured by the annotations). For some domains or use-cases, such a repair engine may suffice. To expand the scope of repairs to semantic fixes, 
such as identifying incorrect function call arity and performing edits to resolve this, we require the
language developer provide domain-specific strategies. There are domain specific strategies (DSS) that are likely to be useful across domains (such as checking call arity) and
need only be instantiated with domain-specific values (such as the function names and associated number of arguments) -- this was our experience with both the Excel and PowerFx domains. Importantly, the language
developer can add DSS incrementally, focusing on implementing a check/fix strategy
for observed user errors. In discussions with partner teams, this data-driven approach to 
DSS development fits well with their traditional workflow.

If the language developer has access to a corpus of well-formed programs in their domain, they may consider training the neural ranker component, which complements the edit-distance-based ranking that takes place in the purely symbolic approach. To train the ranker successfully, the language developer's corpus should contain at least 10s of thousands of well-formed programs. In our experience, these programs can be scraped from online resources (e.g. help forums or repositories) or production resources (e.g. product telemetry). We found that a relatively small model performed well for ranking purposes, so modest GPUs like a K80 work well.

Furthermore, the language developer can also choose to train the neural localizer, to complement the symbolic backtracking implemented by default. If the language developer has pairs of real buggy programs and their corrected versions, as might be available from product telemetry, this data is well suited for training the pointer network. If the paired corpus is relatively small (fewer than 10s of thousands of pairs), the language developer can produce synthetic pairs by introducing errors (uniformly at random) into their corpus of well-formed programs. Additionally, the language developer can train the pointer network initially on these synthetic pairs and \emph{then} fine-tune on their real paired corpus. Similarly to the ranker, we successfully trained the pointer networks for the Excel and PowerFx domains using modest GPU resources.

Note that without training a neural ranker and localizer, the language developer still has the ability of using \lamirage{}, as we can default to their symbolic alternatives. While the resulting repair engine
will fix less programs, as demonstrated by our ablation studies presented in \Cref{sec:ablation},
the purely symbolic system can more easily be deployed in limited-resource platforms such as the browser.
Furthermore, the added value from the neural components may vary across domains -- in practice, it may be
possible to implement post/pre-processing heuristics that cover enough common cases to avoid the need for the 
neural components, when they are cumbersome to deploy.

Finally, the language developer has the option of setting different values for three hyperparameters: the backtracking depth, a maximum local cost, and a maximum global cost -- all of which come with defaults
that work well in our two evaluation domains. The backtracking depth determines how far back edits are made  from the detected error location. The maximum local cost is used to prune out search candidates early that will have a total edit distance higher than the threshold. The maximum global cost is the maximum edit distance threshold allowable for repair candidates returned by the engine.
The appropriate values for each of these hyperparameters depends on the
use case of the language developer. In particular, increasing the backtracking depth and the two cost hyperparameters allows the repair engine to produce candidates with further away edit locations and more edits. This increase in the search space may improve coverage, but will increase search times and may require more investment in the neural ranker to distinguish candidates. Conversely, reducing the values of these hyperparameters may produce fewer repair candidates but return results quickly to the end-user. We expect language developers will have a set of buggy programs to repair as benchmark tasks, and these can be used to pick appropriate hyperparameter values, as is standard in many AI-based systems.

\section{Related Work}
\label{sec:related}
We discuss symbolic and neural error correction, and general program repair and synthesis.

\cheapsubsection{Symbolic Approaches for Error Correction}
Historically, most practical implementations of error correction for mistakes such as
syntax errors have relied on greedy/simple approaches such as panic error recovery~\cite{dragonbook} -- where
the processing system just deletes tokens until it can resume parsing.
However, more complete error recovery strategies do exist~\cite{FischerRecovery, Cerecke2003LocallyLE, Degano1995ComparisonOS,
IkSoon2010, Corchuelo02, Aho72, NicolaeRajasekara14,
SpenkeReliabilityBasedErrorCorrectionSPE1984}. The state-of-the-art symbolic approach
to error recovery, developed by~\citet{ecoop20} and implemented in \grmtools, improves on \citet{Corchuelo02}'s approach
both in completeness (the ability to return minimum edit sets)
and speed (they optimize their implementation).

Like \lamirage, these approaches enumerate repairs as edit operation sequences that allow parsing
to continue whenever an error is encountered.
The key difference with 
\lamirage{} stems from the increased expressiveness of 
its repairs. First, \lamirage{} is syntax-guided but
is not limited to syntax repairs only. \lamirage{} allows language developers to 
add domain-specific strategies, which capture
non-context free properties of well-formed programs, to increase the scope of repairs. This allows 
\lamirage{} repair engines to produce fixes for some semantic mistakes like incorrect function call arity.
Second, \lamirage{} can address non-local errors -- meaning errors that require a repair that is not in the immediate location
where the error is detected -- by performing backtracking.
State-of-the-art symbolic systems such as \grmtools{} do not
perform backtracking as it leads to an explosion in the search space. \lamirage{} mitigates this by combining two ideas. When it backtracks, \lamirage{} still limits editing actions to unreliable tokens (constraining the space of possible candidates). Additionally, it backtracks up to a fixed depth and relies on a neural localizer to identify candidate locations beyond that depth bound.

As both of these ideas increase the size of \lamirage{}'s search space, the system also needs to be more effective at comparing candidates. Tools such as \grmtools{} rely exclusively on minimum edit distance. In contrast, \lamirage{} also employs a neural ranker that can break ties between otherwise equidistant candidates.

\cheapsubsection{Neural approaches for Error Correction}
Program repair systems are increasingly using deep learning to correct syntax and semantic errors 
in general purposes programming languages~\cite{BIFI, DrRepair, DeepFix, Tufano18, Sensibility, SynFix}.
Our evaluation compares to three such methods (\bifi{}, \codex{}, \codexedit{}). Other work in the space includes
DeepFix~\cite{DeepFix}, which fixes syntactic errors in C programs using a sequence-to-sequence approach; 
SynFix~\cite{SynFix}, which fixes Java syntax errors by applying multiple deep learning models in 
sequence; and
TFix~\cite{TFix}, which pretrains a T5 transformer~\cite{raffel2019exploring} on natural language and fine-tunes it on buggy/repaired code snippets mined from Github commits.

A challenge with purely neural methods is the lack of guarantees over outputs.
For example, language models are known to generate plausible but incorrect content, 
known as ``hallucinations''~\cite{guo2021learning}. Such models are also known to introduce
basic errors~\cite{synchromesh}, which complicates repair.
In contrast, \lamirage's candidate repairs are guaranteed to satisfy our approximate grammar $G$ and our constraints $\mathcal{C}$. One way to mitigate this issue with neural models is to increase their training data but availability (at scale) can be challenging in the LC domain.

Moreover, repair tools for general purpose languages, such as 
TFix\footnote{Applying pre-trained TFix to basic Excel formulas failed to generate any relevant repair candidates -- fine-tuning TFix would also require a more detailed Excel analyzer that is not available.}
and SynFix or even non-neural methods such as GetAFix~\cite{GetAFix}, typically
assume a more detailed oracle (static analyzer) that provides detailed diagnostics.
This oracle is not often available for LC domains, in part due to lack of tooling.

\cheapsubsection{General Program Repair and Synthesis}
\citet{monperrus20} reviews the vast literature of automated program repair,
most of which is focused on repair in the context of general purpose programming
languages\cite{CacmOverview, GenProg, PAR, SemFix,Angelix, Prophet,
long2017automatic, GetAFix, apifix} and not low-code domains. In contrast,
\lamirage specifically focuses on last-mile repair in low-code domains, where
oracles such as test suites or static analyzers are not readily available.
The lack of such oracles has influenced the scope and design of
\lamirage{}. Specifically, we focused on repairing errors that require small
changes and that can be detected without substantial additional context.
Intuitively, these error correspond to those that a typical user might post to a
help forum, where an expert low-code user can often understand intent and
provide a fix simply from the posted formula.

As discussed, there are parallels between \lamirage{}'s syntax-guided approach 
and other syntax-guided program synthesis systems~\cite{gulwani2011automating,devlin2017robustfill,polozov2015flashmeta, prose-github}. However, there are key differences. First, we only have the buggy formula
and no additional specification (e.g. input/output examples). Second, a generic string transformation DSL is unlikely to result in valid repairs.
In contrast to synthesis for program transformations~\cite{rolim2017learning, bluepencil}, \lamirage{} does not mine edit patterns from groups of programs but rather relies on the CFG and DSS to induce transformations of the user's buggy formula. Incorporating mined patterns into \lamirage{} as DSS is left as future work.

Prior program repair work has explored the use of machine learning models during search. For example, Prophet~\cite{Prophet} used a log-linear model to rank candidate patches before validating them with a test suite. In contrast to this work, 
\lamirage{} uses two \emph{neural} models during search: a pointer-network-based localizer and a transformer-based ranker. While Prophet computes manually defined features over C patches, \lamirage{} relies on the ability of neural models to capture
high-dimensional patterns without the need for manually defined features.
Prophet produces their candidate patches by applying template-based
rewrites to the originally buggy program. In contrast, \lamirage{}
performs a syntax-guided search as it processes the input buggy
program. Finally, LaMirage performs pruning while searching (based on an approximate edit distance bound) compared to Prophet, which first generates all patches, ranks them, and then validates them using a test suite.

More broadly, combining machine learning
and symbolic methods has been explored by past work
both in the areas of program synthesis generally
and program repair specifically, for example 
~\cite{kalyan2018neural, ellis2018learning, chen2018execution}
and
~\cite{zhu2021syntax, tang2021grammar, yu2019learning}, respectively.
Many of these systems use neural methods to reduce the search
time (by constraining or prioritizing search direction) and
improving the ranking of competing candidates produced.
In this regard, \lamirage{} takes a similar approach. However,
in this work we introduce a novel application of these techniques to 
the task of producing last-mile repairs that fix syntax and
some semantic errors in the low-code domain. Furthermore, we present various insights that show how to effectively combine
these techniques for our particular use case. Specifically, we show
that we can use relatively small models for both 
error localization and ranking in the low-code domain. Both of these can be 
successfully trained with modest amounts of data -- enabled by the smaller model sizes -- collected from a combination
of help forums and product telemetry. In the case of the neural localizer, we show that
using a fallback strategy, where we employ the neural model
only when symbolic tracking fails to produce a candidate, lets us
exploit the additional coverage
of the neural localizer without blowing up the search space.
In the case of the neural ranker, we show that applying a standard causal language modeling objective fails to work due to the skew in unreliable token sequences -- we introduced instead a balanced training set and a classification task that can be iteratively applied to produce appropriate likelihood predictions.

\cheapsubsection{Neurosymbolic Methods for Programming Tasks}
Prior work in the programming languages and software engineering communities has
explored the use of combining neural and symbolic methods to assist developers with 
programming tasks.

Synchromesh~\cite{synchromesh} improves the extent to which large language models, such as Codex, 
can generate syntactically and semantically valid code from natural language utterances.
To do so, Synchromesh biases the decoding process towards tokens that would produce valid
completions. These tokens are derived from the associated grammar and domain constraints (such as table schemas for SQL). 
Additionally, Synchromesh introduces target similarity tuning -- 
a method for picking few shots for the prompt that are expected to have a similar program structure to the intended
program (as specified via the natural language utterance).

\citet{multimodalcbs} note that LLMs often fail to generate the desired program directly from natural language descriptions, but the programs that they do generate often contain most or all of the components that should 
appear in the desired program. Thus, their approach consists of generating candidate programs from the natural language utterance using an LLM, and then mining code fragments from these as components, and performing component-based synthesis over these mined components (and a DSL) to satisfy the input/output examples
provided. In particular, not only are the components from the LLM useful for initializing the search, but other features of the programs from the LLMs also help for pruning and guiding the beam search, as well as ranking the synthesized candidates to learn programs more efficiently and from fewer examples. They show that such an approach performs well in the domains of regular expressions and CSS selectors.

\citet{verbruggen2021semantic} combine LLMs into a traditional inductive synthesizer by identifying subproblems that cannot be solved through syntactic transformations defined in the base DSL and can be handed off to a LLM to produce a solution. By incorporating an LLM into a symbolic synthesizer, their approach can support semantic transformations of inputs, such as returning the currency symbol given a country name, without the need to manually define such operators.

Similarly to this line of work, \lamirage{} combines both symbolic and neural methods. In particular, \lamirage{}
uses a neural localizer and ranker, and integrates these into a symbolic (syntax-guided) framework for
processing buggy input programs and generating candidate repairs. The contributions of \lamirage{} are
to apply these ideas to a new task (last-mile repair in the low-code domain) and to arrive at the
specific design that works well in practice (as shown by our evaluation).

\section{Conclusion}\label{sec:conc}

We presented \lamirage, a last-mile repair-engine generator
for programs written in low-code (LC) formula languages.
\lamirage targets ``last-mile repairs'', where
the formula is almost correct and has a few subtle errors.
We designed \lamirage to combine the advantages of symbolic 
and neural techniques. We evaluated
\lamirage on real Excel and PowerFx formulas. 
Our results showed that \lamirage outperforms state-of-the-art symbolic and neural
techniques in both domains. We carried out ablation studies on the design decisions in \lamirage.
We discussed the useability of our
framework and design considerations, motivated
by our ongoing partnership with industry teams to 
integrate our repair engines into leading 
LC platforms. Finally, we are releasing our two benchmark sets for future work in low-code domains.

\bibliography{references}

\clearpage
\appendix
\section{Appendix}
\section{Neural Localizer}
We use a publicly available implementation of Pointer Network for Neural Error Localizer:  \url{https://github.com/ast0414/pointer-networks-pytorch}. 
We use cross entropy loss and Adam optimizer for training. We train our model for 26 hours and 7 hours for Excel and Powerapps respectively.

\subsection{Training Data} 
We train pointer network independently for Excel with 480572 synthetically generated well formed-buggy formula pairs and Powerfx with 60802 synthetically generated well formed-buggy formula pairs.
\\
\textbf{Vocabulary size: }The vocabulary size for Excel was 12344 and for Powerfx it was 86224.
\\
\textbf{Data Preprocessing: } We anonymize cell references, string literals and numbers for Excel. We anonymize string literals and numbers for PowerFx.

\begin{table}[h]
\begin{tabular}{l|r}
\textbf{Hyperparameters} & Values                  \\ \hline
Batch Size               & 256                     \\
Embedding Dim            & 256                     \\
Encoder Layers           & 4                       \\
Hidden Size              & 256                     \\
Learning Rate            & 0.0001                  \\
Loss Reduction           & sum                     \\
Early Stopping Patience  & 5                       \\
Number of Epochs         & 100
\end{tabular}
\caption{Hyperparameters for trained Pointer Network for Excel and Powerapps}
\end{table}

\section{Detailed \codex{} and \codexedit{} Tables}

\begin{table}[!h]
\begin{tabular}{lrrr}
\toprule
Approach &  Top-1 &  Top-3 &  Top-5 \\
\midrule
\codex{}(t=0.0) &    110 &    115 &  115 \\
\codex{}(t=0.1) &    110 &    120 &  120 \\
\codex{}(t=0.3) &    111 &    139 &  139 \\
\codex{}(t=0.4) &    110 &    148 &  149 \\
\codex{}(t=0.5) &    108 &    152 &  154 \\
\codex{}(t=0.7) &    111 &    156 &  160 \\
\codexedit{}(t=0.0) &    145 &    146 &  146 \\
\codexedit{}(t=0.1) &    147 &    153 &  153 \\
\codexedit{}(t=0.3) &    146 &    163 &  163 \\
\codexedit{}(t=0.4) &    142 &    163 &  165 \\
\codexedit{}(t=0.7) &    118 &    158 &  165 \\
\codexedit{}(t=0.5) &    135 &    163 &  163 \\
\bottomrule
\end{tabular}
\caption{Detailed Excel Results}
\end{table}

\begin{table}[!h]
\begin{tabular}{llrrr}
\toprule
Approach &  Top-1 &  Top-3 &  Top-5 \\
\midrule
\codex{}(t=0.0) &     79 &     83 &   83 \\
\codex{}(t=0.1) &     81 &     92 &   92 \\
\codex{}(t=0.3) &     84 &    112 &  117 \\
\codex{}(t=0.4) &     86 &    114 &  130 \\
\codex{}(t=0.5) &     86 &    114 &  127 \\
\codex{}(t=0.7) &     85 &    117 &  132 \\
\codexedit{}(t=0.0) &    106 &    107 &  107 \\
\codexedit{}(t=0.1) &    105 &    120 &  120 \\
\codexedit{}(t=0.3) &    106 &    129 &  131 \\
\codexedit{}(t=0.4) &    105 &    137 &  138 \\
\codexedit{}(t=0.5) &     94 &    134 &  140 \\
\codexedit{}(t=0.7) &     79 &    122 &  137 \\
\bottomrule
\end{tabular}
\caption{Detailed PowerFx Results}
\end{table}

\end{document}